\UseRawInputEncoding
\documentclass[journal]{IEEEtran}
\usepackage{cite}
\usepackage[cmex10]{amsmath}
\usepackage{amssymb}
\usepackage{amsthm}
\usepackage{mathrsfs}
\usepackage{bm}
\usepackage[mathscr]{eucal}
\usepackage{amssymb,amsmath,amsthm,amsfonts,latexsym}
\usepackage{amsmath,graphicx,bm,xcolor,url}
\usepackage{graphicx}
\usepackage{latexsym}
\usepackage{CJK}
\usepackage{indentfirst}
\usepackage{geometry}
\usepackage{psfrag}
\usepackage{setspace}
\usepackage{algorithmic}
\usepackage{algorithmic, cite}
\usepackage{algorithm}
\usepackage{array}
\usepackage{mdwmath}
\usepackage{mdwtab}
\usepackage{eqparbox}
\usepackage{url}
\usepackage{epstopdf}
\usepackage{epsfig,epsf,psfrag}
\usepackage{verbatim}
\usepackage{textcomp}
\usepackage{psfrag} 

\usepackage{subcaption}

\theoremstyle{plain}

\theoremstyle{remark}

\theoremstyle{plain}

\theoremstyle{remark}

\theoremstyle{plain}

\theoremstyle{remark}

\theoremstyle{remark}

\theoremstyle{remark}

\theoremstyle{remark}

\theoremstyle{remark}

\theoremstyle{remark}

\catcode`~=11 \def\UrlSpecials{\do\~{\kern -.15em\lower .7ex\hbox{~}\kern .04em}} \catcode`~=13

\allowdisplaybreaks[4]

\newcommand{\calD}{\mathcal{D}}

\newcommand{\bE}{\mathbf{E}}

\DeclareMathAlphabet{\mathbsf}{OT1}{cmss}{bx}{n}
\DeclareMathAlphabet{\mathssf}{OT1}{cmss}{m}{sl}

\DeclareSymbolFont{bsfletters}{OT1}{cmss}{bx}{n}
\DeclareSymbolFont{ssfletters}{OT1}{cmss}{m}{n}
\DeclareMathSymbol{\bsfGamma}{0}{bsfletters}{'000}
\DeclareMathSymbol{\ssfGamma}{0}{ssfletters}{'000}
\DeclareMathSymbol{\bsfDelta}{0}{bsfletters}{'001}
\DeclareMathSymbol{\ssfDelta}{0}{ssfletters}{'001}
\DeclareMathSymbol{\bsfTheta}{0}{bsfletters}{'002}
\DeclareMathSymbol{\ssfTheta}{0}{ssfletters}{'002}
\DeclareMathSymbol{\bsfLambda}{0}{bsfletters}{'003}
\DeclareMathSymbol{\ssfLambda}{0}{ssfletters}{'003}
\DeclareMathSymbol{\bsfXi}{0}{bsfletters}{'004}
\DeclareMathSymbol{\ssfXi}{0}{ssfletters}{'004}
\DeclareMathSymbol{\bsfPi}{0}{bsfletters}{'005}
\DeclareMathSymbol{\ssfPi}{0}{ssfletters}{'005}
\DeclareMathSymbol{\bsfSigma}{0}{bsfletters}{'006}
\DeclareMathSymbol{\ssfSigma}{0}{ssfletters}{'006}
\DeclareMathSymbol{\bsfUpsilon}{0}{bsfletters}{'007}
\DeclareMathSymbol{\ssfUpsilon}{0}{ssfletters}{'007}
\DeclareMathSymbol{\bsfPhi}{0}{bsfletters}{'010}
\DeclareMathSymbol{\ssfPhi}{0}{ssfletters}{'010}
\DeclareMathSymbol{\bsfPsi}{0}{bsfletters}{'011}
\DeclareMathSymbol{\ssfPsi}{0}{ssfletters}{'011}
\DeclareMathSymbol{\bsfOmega}{0}{bsfletters}{'012}
\DeclareMathSymbol{\ssfOmega}{0}{ssfletters}{'012}

\def\norm#1{\left\| #1 \right\|}
\def\norm2#1{\left\| #1 \right\|_2}
\def\norm22#1{\left\| #1 \right\|_2^2}

\newcommand{\qednew}{\nobreak \ifvmode \relax \else
      \ifdim\lastskip<1.5em \hskip-\lastskip
      \hskip1.5em plus0em minus0.5em \fi \nobreak
      \vrule height0.75em width0.5em depth0.25em\fi}
\usepackage{cite}
\usepackage[cmex10]{amsmath}%

\usepackage{mathrsfs}
\usepackage{bm}
\usepackage[mathscr]{eucal}
\usepackage{amssymb,amsmath,amsthm,amsfonts,latexsym}
\usepackage{amsmath,graphicx,bm,xcolor,url}
\usepackage{graphicx}
\usepackage{latexsym}
\usepackage{CJK}
\usepackage{indentfirst}
\usepackage{geometry}
\usepackage{psfrag}
\usepackage{setspace}
\usepackage{algorithmic}
\usepackage{algorithmic, cite}
\usepackage{algorithm}
\usepackage{array}
\usepackage{mdwmath}
\usepackage{mdwtab}
\usepackage{eqparbox}
\usepackage{url}
\usepackage{epstopdf}
\usepackage{epsfig,epsf,psfrag}
\usepackage{verbatim}
\usepackage{textcomp}
\usepackage{float}
\usepackage{cite}
\usepackage{times,amsmath,color,amssymb,graphicx,epsfig,multirow,float,algorithm,algorithmic,array}
\usepackage{stackengine}
\usepackage{graphicx,float}

\newcommand\xrowht[2][0]{\addstackgap[.5\dimexpr#2\relax]{\vphantom{#1}}}
\title{Low-Complexity Joint Azimuth-Range-Velocity Estimation for Integrated Sensing and Communication with OFDM Waveform}
\author{Jun Zhang, Gang~Yang, \emph{Member, IEEE}, Qibin Ye, Yixuan Huang, and Su~Hu, \emph{Member, IEEE}
\thanks{J. Zhang, Q. Ye, Y. Huang and S. Hu are with National Key Laboratory of Wireless Communications, University of Electronic Science and Technology of
China, Chengdu 611731, China (e-mails: \{202322220221,202111220623\}@std.uestc.edu.cn;\{huangyx,husu\}@ues\\
tc.edu.cn).
}
\thanks{G. Yang is with the Shenzhen Institute for Advanced Study, and the National Key Laboratory of Wireless Communications, University of
Electronic Science and Technology of China, Chengdu 611731,China (e-mail: yanggang@uestc.edu.cn) (Corresponding author: Gang Yang).
}
}

\begin{document}
\maketitle 
\begin{abstract}
Integrated sensing and communication (ISAC) is a main application scenario of the sixth-generation mobile communication systems. Due to the fast-growing number of antennas and subcarriers in cellular systems, the computational complexity of joint azimuth-range-velocity estimation (JARVE) in ISAC systems is extremely high. This paper studies the JARVE problem for a monostatic ISAC system with orthogonal frequency division multiplexing (OFDM) waveform, in which a base station receives the echos of its transmitted cellular OFDM signals to sense multiple targets. The Cram\'{e}r-Rao bounds are first derived for JARVE. A low-complexity algorithm is further designed for super-resolution JARVE, which utilizes the proposed iterative subspace update scheme and Levenberg-Marquardt optimization method to replace the exhaustive search of spatial spectrum in multiple-signal-classification (MUSIC) algorithm. Finally, with the practical parameters of 5G New Radio, simulation results verify that the proposed algorithm can reduce the computational complexity by three orders of magnitude and two orders of magnitude compared to the existing three-dimensional MUSIC algorithm and estimation-of-signal-parameters-using-rotational-invariance-techniques (ESPRIT) algorithm, respectively, and also improve the estimation performance.
\end{abstract}


\vspace{-0.25cm}
\section{Introduction}\label{introduction}

The integrated sensing and communication (ISAC) that shares the spectrum and infrastructure resources for communication and sensing, has been recognized as a key technology for the sixth-generation (6G) mobile communication systems~\cite{ITU6GFramework2023}. ISAC has tremendous application areas like internet-of-vehicles (IoV) networks~\cite{application12}, unmanned-aerial-vehicle (UAV) networks~\cite{application2} and military networks~\cite{application3}. The signal waveform is a basis for designing efficient ISAC systems~\cite{ISACDesign,wave3,wave4}. The main ISAC waveform candidates include the frequency modulated continuous waveform (FMCW), the orthogonal-time-frequency-space (OTFS) waveform and the orthogonal-frequency-division-multiplexing (OFDM) waveform. The FMCW can estimate both the range and velocity parameters precisely due to its broad bandwidth and long time duration~\cite{FMCW1,FMCW2,FMCW3}, but suffers from low spectrum efficiency for wireless communications~\cite{wave4}. The OTFS waveform is preferable for high-speed application scenarios benefiting from its robustness to fast time-varying channels~\cite{otfs1}, but requires complex signal processing and is still in its infancy~\cite{otfs2}. In contrast, the OFDM waveform, which achieves high spectrum efficiency in widely-used wireless systems like cellular networks and wireless fidelity networks, has great potential to achieve precise estimation of multidimensional parameters, as it is broadband continuous-time signals~\cite{OFDM1,OFDM-huang,OFDM2,OFDM3,OFDM4}. Therefore, this paper focuses on the ISAC systems with OFDM waveforms. 

In particular, the parameter estimation algorithm is the most critical design content for precise sensing in ISAC systems. In 6G era, the millimeter-wave/Terahertz  broadband signals and the ultra-large-scale multi-input-multi-output (MIMO) antenna array will pave the way for high-precision parameter estimation~\cite{6G-trends,6G-Thz}, but also result into extremely high or even unaffordable computational complexity, especially for joint estimation of multidimensional parameters. Thus, it is desirable to achieve optimal trade-offs between estimation performance and computational complexity. 

The conventional algorithms for jointly estimating multidimensional parameters are mainly based on the Discrete-Fourier-Transform (DFT) operation or the minimum-variance-distortionless-response criterion. The DFT-based algorithms utilize the conventional beamforming in time, frequency and space domains to extract the phase rotation resulted from each parameter, which can be efficiently implemented by Fast-Fourier-Transform circuit~\cite{DFT,3D-DFT}. To improve the estimation performance, the Capon algorithm was first proposed in~\cite{capon} for one-dimensional azimuth estimation, which utilizes adaptive beamforming to minimize the variance-distortionless-response, and further extended to the two-dimensional (2D) Capon algorithm for joint range-velocity estimation~\cite{2dcapon}. However, the aforementioned conventional algorithms have limited estimation resolution given by the Rayleigh limit, which is inversely proportional to the sensing resources including the time duration, bandwidth, and array aperture\cite{3D-DFT}.

To break through the Rayleigh limit and thus achieve super-resolution estimation, two subspace-based algorithms, namely, the multiple-signal-classification (MUSIC) algorithm and the estimation-of-signal-parameters-using-rotational-invariance-techniques (ESPRIT) algorithm, were proposed for one-dimensional parameter estimation in~\cite{doa-MUSIC} and~\cite{doa-ESPRIT}, respectively, and extended for multidimensional parameter estimation. The multidimensional MUSIC estimation algorithms~\cite{2D1,2D2,2D3,2D4,3D-MUSIC} first separate the signal subspace and the noise subspace, then perform multidimensional spatial spectrum search based on the orthogonality between the signal subspace and noise subspace. The three-dimensional (3D) MUSIC algorithm~\cite{3D-MUSIC} realized the super-resolution JARVE and auto-pairing of parameters. However, due to its exhaustive on-grid search, the computational complexity of the 3D-MUSIC algorithm increases steeply with the number of grid points decided by estimation accuracy. In contrast, the 3D-ESPRIT algorithm proposed in~\cite{3D-ESPRIT-hu,3D-ESPRIT1}, performs off-grid JARVE by exploiting the subspace rotation invariance characteristic of the 3D Vandermonde manifold matrices. The complexity of the 3D-ESPRIT algorithm is irrelevant to the estimation accuracy, but proportional to the cubic of the signal observation sizes in the space-frequency-time domains. Since the sizes of signal observations in future ISAC systems are fast growing with the large number of antennas and subcarriers, the complexity of the 3D-ESPRIT algorithm is still high. Hence, it is desirable to design low-complexity multidimensional parameter estimation algorithms.



In this paper, we study the low-complexity super-resolution JARVE problem in an ISAC system with OFDM waveform. The main contributions are summarized as follows.
\begin{itemize}
\item We establish the signal model for an OFDM-based ISAC system in which a base station (BS) utilizes its transmitted cellular OFDM signals to sense multiple targets, and obtain the Cram\'{e}r-Rao bounds (CRBs) for the joint azimuth-range-velocity estimation by deriving the Fisher information matrix. Specifically, for single-target sensing, the close-form CRB of each-dimension parameter is inversely proportional to the signal-to-noise ratio (SNR), the number of snapshots and the cubic of degree of freedom in the corresponding dimension. 

\item For low-complexity JARVE, we propose a parallel iterative-subspace-updating 2D-MUSIC (PI-2DMUSIC) algorithm that consists of the spatial smoothing step, the 3D-parameter initialization step, and the ISU-2DMUSIC parallel estimation as well as 3D-parameter pairing step. For the {\textit{first}} step, the spatial smoothing operation is performed for the decoherence of signal observation components from multiple targets. For the {\textit{second}} step, a low-complexity initialization scheme is designed for all the 3D parameters. For the {\textit{third}} step, the 3D matrix of signal observations in space-frequency-time domains are sliced into a series of 2D matrices from one dimension, and the proposed ISU-2DMUSIC algorithm utilizes the combined covariance matrices for the vectorized 2D matrices to estimate the 2D parameters. Specifically, the ISU-2DMUSIC algorithm includes two-layer iterations. In each outer-layer iteration, each estimated target's rotation vector is projected into the noise subspace to eliminate the interference of echo signals from the estimated targets; while in each inner-layer iteration, the efficient Levenberg-Marquardt optimization method is utilized to replace the exhaustive search process of spatial spectrum in conventional MUSIC method. The 2D-MUSIC estimation step is performed in parallel for all 3D-to-2D slicing from each dimension. The estimated 2D parameters are then matched to the 3D-parameter pairs for all targets by  following the minimum-distance criterion. 
 


\item We analyze the complexity of the proposed PI-2DMUSIC algorithm and show its significant complexity reduction compared to the mainstream super-resolution JARVE algorithms. The complexity of the PI-2DMUSIC algorithm is ${\rm O}\left( {\tilde Z_1^3\tilde Z_2^3} \right)+{\rm O}\left( {\tilde Z_1^2\tilde Z_2^2} S_3 \right)$, where ${{\tilde Z}_1} > {{\tilde Z}_2}>{{\tilde Z}_3}$ denote a decreasing sort of the space-domain size $\tilde L$, frequency-domain size $\tilde N$ and time-domain size $\tilde M$ of the spatially-smoothed sub-signal observation, $S_3=Z_3-\tilde Z_3+1$, and $Z_3$ is the size of the original signal observation in the domain corresponding to $\tilde Z_3$. In contrast, the complexity of the 3D-ESPRIT algorithm~\cite{3D-ESPRIT-hu} is ${\rm O}\left( {{{ {\tilde L^3\tilde N^3 \tilde M^3}}}} \right)+{\rm O}\left( {{{ {\tilde L^2 \tilde N^2 \tilde M^2}}S}} \right)$, where the number of spatially-smoothed snapshots $S=(L-\tilde L+1)(N-\tilde N+1)(M-\tilde M+1)$, with $L$, $N$ and $M$ representing the number of antennas, subcarriers and echo symbols, respectively; while the complexity of the 3D-MUSIC algorithm~\cite{3D-MUSIC} is ${\rm O}\left( {{{\tilde L^2 \tilde N^2 \tilde M^2}}}{G_\theta }{G_r}{G_v} \right)$, with ${G_\theta },{G_r},{G_v}$ representing the grid number for searching azimuth, range and velocity, respectively. Since it typically holds that ${\tilde Z_1\tilde Z_2} \ll {\tilde L\tilde N\tilde M} \ll {G_\theta }{G_r}{G_v}$, and $S_3 \ll S \ll {G_\theta }{G_r}{G_v}$, the complexity of the proposed algorithm is much lower than the two mainstream super-resolution JARVE algorithms.
\item Simulation results are given to verify the computational complexity and estimation performance of the proposed algorithm. On the one hand, the proposed PI-2DMUSIC algorithm can achieve computational complexity reduction by three orders of magnitude and two orders of magnitude compared to the super-resolution 3D-MUSIC algorithm and 3D-ESPRIT algorithm, respectively, as well as the estimation performance improvement. Specifically, for the case of 32 receive antennas, 128 subcarriers and 80 OFDM symbols, the proposed PI-2DMUSIC algorithm reduces the computational complexity by 2242 times and 149 times, compared to the 3D-MUSIC algorithm and 3D-ESPRIT algorithm, respectively. On the other hand, compared to the traditional 3D-DFT algorithm with limited resolution, the proposed PI-2DMUSIC algorithm reduces the root of minimum square error (RMSE) by at least two orders of magnitude, at the cost of slight complexity increase. More importantly, the proposed algorithm's complexity reduction and estimation performance improvement become more obvious for larger SNR and more sensing resources in the space-frequency-time domains. 



\end{itemize}

This paper is organized as follows: Section~\ref{SystemModel} presents the system model of ISAC with OFDM waveform. Section~\ref{sec:CRB} derives the CRBs for JARVE. Section~\ref{sec:Proposed Algorithm} proposes a super-resolution JARVE algorithm and analyzes its complexity. Section~\ref{sec:simulation} gives simulation results to verify the performance and complexity of proposed algorithm. 
 
{\emph{Notations}}:  Throughout this paper, $[{\bf A}]^T$, $[{\bf A}]^H$, $[{\bf A}]^{-1}$, $[{\bf A}]^\dag$, ${\|\bf A\|}_2$, ${\mathop{\rm Re}\nolimits}\{\bf A\}$, ${\mathop{\rm Im}\nolimits}\{\bf A\}$, ${\text {Tr}}{({\bf A})}$ and ${\mathbb E}{({\bf A})}$ denote the transpose, hermitian, inverse, Moore-Penrose pseudoinverse, Frobenius norm, real part, imaginary part, trace and mathematical expectation of a matrix $\bf A$, respectively. 
$\angle{a}$ denotes the phase of a complex number $a$. ${{\bf{I}}_u}$ denotes an $u$-dimensional unitary matrix, ${{\bf{0}}_{u\times v}}$ denotes an all-0 matrix with  $u$ rows and  $v$ columns. ${\text {diag}}\left\{ \cdot \right\}$ and ${\text {blkdiag}}\left\{ \cdot \right\}$ denote the diagonal matrix and the block diagonal matrix, respectively.  $\odot$ and $\otimes$ denote the Khatri-Rao product and the Kronecker product, respectively. $\mathcal{CN}(\mu,{ \sigma ^2})$ denotes the circularly symmetric complex Gaussian (CSCG) distribution with mean $\mu$ and variance $\sigma^2$. 



\section{System Model} \label{SystemModel}

In this section, we present the system description and signal model for ISAC with OFDM waveform.


\subsection{System Description}\label{subsec:Sending Signal_Model}

\begin{figure}[t]
\vspace{-0.1cm}
\centering
\includegraphics[width=.99\columnwidth] {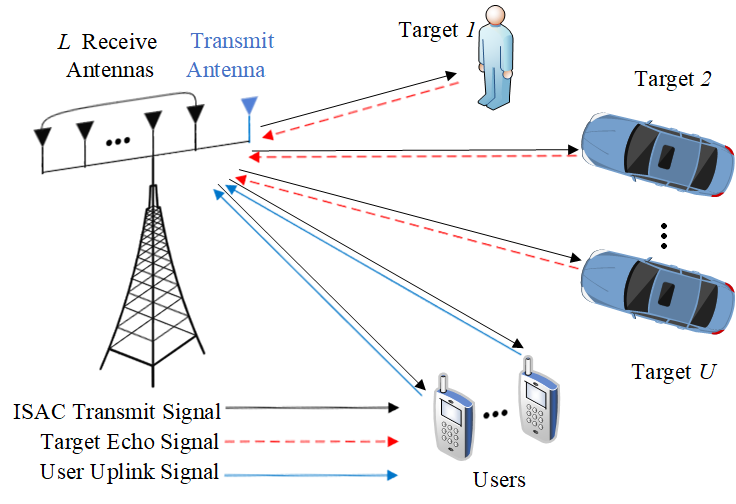}
\caption{System Model for ISAC over OFDM waveform.}
\label{fig:Fig1}
\vspace{-0.3cm}
\end{figure}

As demonstrated in Fig.~\ref{fig:Fig1}, the considered ISAC system consists of a base station (BS) with ISAC function, multiple celluar users, and $U$ sensing targets. The BS uses one antenna to send downlink sensing signal, and uses other $L$ antennas to receive the echo signals reflected from the targets such as vehicles and roadblocks. The BS processes the echo signals to obtain the JARVE and realize the sensing function. In practice, the ISAC system can reuse the downlink data signal or reference signal with OFDM waveform as the sensing signal. The typical application scenario of the studied ISAC system includes the cellular network enabled IoV networks, pedestrian sensing, etc. 

\subsection{Signal Model}\label{subsec:Signal_Model}
Without loss of generality, suppose that the transmitted OFDM sensing signal has $N$ subcarriers, and denote the subcarrier bandwidth by $\Delta f$. Let the carrier frequency be ${f_{\rm{c}}}$, and it typically holds that ${f_{\rm{c}}} \gg N\Delta f$. The transmitted time-domain OFDM sensing signal in the $m$-th symbol period can be expressed as
\begin{align}\label{eq:1}
{x_m}(t) = \sum\limits_{n = 0}^{N - 1} {{S_{m,n}}} u \left( {t - m\bar T} \right){e^{j2\pi ({f_{\rm{c}}} + n\Delta f)t}},
\end{align}
for $m = 0, \ldots,M-1$, and $m\bar T - {T_{{\rm{cp}}}} \le t \le m\bar T + T$, where the period of OFDM data is $T= 1/ \Delta f$, the period of the cyclic prefix (CP) is $T_{\rm{cp}}$, the overall OFDM symbol period is $\bar T= T +T_{{\rm{cp}}}$, ${S_{m,n}}$ is the frequency-domain data symbol transmitted in the $m$-th OFDM symbol and the $n$-th subcarrier, and the pulse signal $u (t)$ is given by
\begin{align}\label{eq:2}
u (t) = \left\{ \begin{array}{l}
1,{\rm{      }}t \in [ - {T_{{\rm{cp}}}},T]\\
0,{\rm{    else.}}
\end{array} \right.
\end{align}

\begin{figure}[htbp]
\vspace{-0.1cm}
\centering
\includegraphics[width=.99\columnwidth] {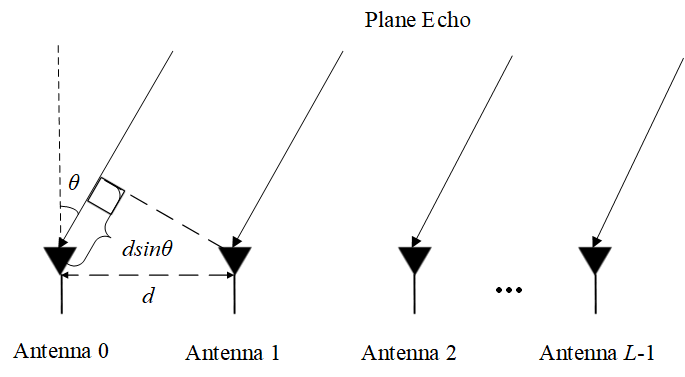}
\caption{Receive Uniform Line Array Model.}
\label{fig:Fig2}
\vspace{-0.3cm}
\end{figure}

As shown in Fig.~\ref{fig:Fig2}, the $L$ receive antennas of the BS are arrayed in an uniform line array (ULA) with the antenna spacing $d$, and the receive antenna with index $0$ is taken as the reference antenna, without loss of generality. The transmitted signal $x_m(t)$ is reflected by the $i$-th target with directions of arrival (DOA) $\theta_i$, range $r_i$, and velocity ${v_i}$, for $i = 1,2, \ldots,U$. Denote the light velocity and the carrier wavelength as $c$ and $\lambda=c/{f_{\rm{c}}}$, respectively. Denote the wavelength of $n$-th subcarrier as $\lambda_n$, for $n = 0,1, \ldots,{N} - 1$, and the corresponding maximum Doppler frequency shift is $f_{{\rm d},i,n}\approx f_{{\rm d},i}=
2{v_i}f_{\rm{c}} /c$. 

From \eqref{eq:1}, the echo signal $y_{l,m} (t)$ received by the $l$-th receive antenna for the $m$-th OFDM symbol is given by
\begin{align}\label{eq:yt_time_cont}
{y_{l,m}}(t) &= \sum\limits_{n = 0}^{N-1} \sum\limits_{i=1}^U   y_{i,l,n,m} (t)\nonumber\\
 &= \sum\limits_{n = 0}^{N-1} \sum\limits_{i=1}^U \eta _i {{S_{m,n}}} {e^{j2\pi ({f_{\rm{c}}} + n\Delta f)t}} e^{j \Phi_{i,l,n,m}} e^{-j 2\pi 2 r_i /\lambda} \nonumber \\
 &\quad + {w}_{l,m}(t),
\end{align}
where $y_{i,l,n,m}(t)$ is the echo signal in the $n$-th subcarrier reflected by the $i$-th target, $\eta_i$ denotes the time-domain target backscatter coefficient of the $i$-th target, the term $2\pi r_i /\lambda$ denotes the propagation phase for the reference echo signal $y_{i,0,0,0}(t)$, the phase rotation $\Phi_{i,l,n,m}$ of $y_{i,l,n,m}(t)$ compared to the phase of $y_{i,0,0,0}(t)$, and ${w}_{l,m}(t) \sim \mathcal{CN}(0,{\sigma ^2})$ denotes the time-domain CSCG noise at the BS.

Specifically, since the inequality $f_{{\rm d},i,n}\bar T \ll  1$ for low-or-medium speed application scenarios like IoV, the phase rotation within
one OFDM symbol can be approximated as a constant~\cite{2dcapon}\cite{2D1}, the phase rotation $\Phi_{i,l,n,m}$ can be expanded as follows
\begin{align}\label{eq:total_delay}
&\Phi _{i,l,n,m} =2\pi \left( {ld\sin {\theta _i} - 2{r_i}} \right)/{\lambda _{n}}+ 2\pi f_{{\rm d},i}m{\bar T} +2\pi 2 {r_i}/\lambda \nonumber\\
&= 2\pi \left( {{f_{\rm{c}}} + n\Delta f} \right)(ld\sin {\theta _i} - 2{r_i})/c + 2\pi f_{{\rm d},i}m{\bar T} \nonumber\\
&\quad  + 2\pi f_{\rm{c}} 2 {r_i}/c \nonumber\\
&\approx 2\pi ld\sin {\theta _i}/\lambda- 2\pi n\Delta f2{r_i}/c   +2\pi m2{v_i}{f_{\rm{c}}}\bar T/c, 
\end{align}
where the approximation in~\eqref{eq:total_delay} comes from the practical facts that ${Ld \ll r_i}$ for far-field propagation, and $N\Delta f \ll {f_{\rm{c}}}$ for OFDM waveform\cite{3D-ML}\cite{3D-cb}. 

From~\eqref{eq:total_delay}, the resulting overall phase rotation $\Phi_{i,l,n,m}$ consists of the three decoupled phase-rotation components, i.e, the phase rotation $2\pi ld\sin {\theta _i}/\lambda$ from the space domain, the phase rotation $-2\pi n\Delta f2{r_i}/c$ from the frequency domain, and the phase rotation $2\pi m2{v_i}{f_{\rm{c}}}\bar T/c$ from the time domain. Specifically, the phase-rotation components for adjacent antennas, subcarriers and symbols are respectively defined as 
\begin{align}\label{eq:phase}
{\Phi _{{\theta _i}}} &\buildrel \Delta \over = 2\pi d\sin {\theta _i}/\lambda,\nonumber\\
{\Phi _{{r_i}}} &\buildrel \Delta \over = -2\pi 2{r_i}\Delta f/c,\nonumber\\
{\Phi _{{v_i}}} &\buildrel \Delta \over = 2\pi 2{v_i}{f_{\rm{c}}}\bar T/c.
\end{align}

After down conversion, discrete-time sampling with interval $\Delta t = {{T} \mathord{\left/ {\vphantom {{T} N}} \right.
 \kern-\nulldelimiterspace} N}$, and CP removal to~\eqref{eq:yt_time_cont}, the $k$-th sample of the time-domain echo signal received by the $l$-th antenna in the $m$-th OFDM symbol, for $k = 0, \ldots,{{N-1}}$, can be written as
\begin{align}\label{eq:yt}
{\tilde y_{l,m}}\left[ k \right] &\buildrel \Delta \over =  \sum\limits_{n = 0}^{N-1} \sum\limits_{i = 1}^U {\tilde y}_{i,l,n,m} (mT + k\Delta t)\nonumber\\
  &= \sum\limits_{n = 0}^{N-1} {\sum\limits_{i = 1}^U { \eta _i{S_{m,n}} {{e^{jl{\Phi _{{\theta _i}}}}}{e^{ jn{\Phi _{{r_i}}}}}{e^{jm{\Phi _{{v_i}}}}}} } }  \cdot  \nonumber\\
&\quad \qquad \qquad {e^{ -j2\pi 2{r_i}/\lambda}}{e^{j2\pi nk /N}} + {\tilde w}_{l,m} \left[ k \right],
\end{align}
where the CSCG noise term ${\tilde w}_{l,m}\left[ k \right] \buildrel \Delta \over = w_{l,m}(mT + k\Delta t){e^{-j2\pi {f_{\rm{c}}}(mT + k\Delta t)}}$.


By performing DFT to~\eqref{eq:yt}, each frequency-domain echo sample at the $n$-th subcarrier can be expressed as
\begin{align}\label{eq:3}
{Y_{l,m}}[n]& = S_{m,n} \sum\limits_{i = 1}^U {{\alpha _i}{e^{j l{\Phi _{{\theta _i}}}}}{e^{j n{\Phi _{{r_i}}}}}e^{  j m{\Phi _{{v_i}}}}} +{{\bar w_{l,m}}[n],}
\end{align}
where $\alpha _i \buildrel \Delta \over = {\sqrt N }{\eta _i}{e^{-j2\pi {2{r_i}/\lambda}}}$ denotes the frequency-domain target backscatter coefficient of the $i$-th target, ${{\bar w}_{l,m}}[n] = \frac{1}{{\sqrt N }}\sum\nolimits_{k = 0}^{N - 1} {\tilde w_{l,m}\left[ k \right]} {e^{ - j2\pi nk/N}} \sim \mathcal{CN}(0,{ \sigma ^2})$ denotes the frequency-domain CSCG noise. 

Dividing each frequency-domain echo sample in~\eqref{eq:3} by the known symbol $S_{m,n}$ yields the signal observation $Z_{l,m}[n] ={Y_{l,m}}[n]/S_{m,n}$ as follow
\begin{align}\label{eq:cijk}
{Z_{l,m}}[n] =\sum\limits_{i = 1}^U {{\alpha _i}{e^{j l{\Phi _{{\theta _i}}}}}{e^{j n{\Phi _{{r_i}}}}}{e^{  j m{\Phi _{{v_i}}}}} + {\omega  _{l,m}}[n]},
\end{align}
where the noise term ${\omega  _{l,m}}[n]={\bar w_{l,m}}[n]/S_{m,n}$. For any phase-shift-keying (PSK) modulated downlink data symbol or constant-modulus sequences generated reference signal symbol $S_{m,n}$, the noise ${\omega  _{l,m}}[n]$ still follows the distribution $\mathcal{CN}(0,{ \sigma ^2})$ (See Lemma 1 in~\cite{3D-ESPRIT-hu} with proof details in Appendix A therein.), which is taken as an assumption for the latter CRB derivation and JARVE algorithm design in this paper.

From \eqref{eq:phase}, the rotation vectors from the antenna, subcarrier and symbol dimension, which carry information of azimuth, range and velocity, respectively, and overall rotation vector, are defined as
\begin{align}\label{eq:a123}
&{{\bf{a}}_{ \theta_i} } \buildrel \Delta \over= {[ {1,{e^{j {\Phi _{{\theta _i}}}}}, \ldots,{e^{j (L - 1){\Phi _{{\theta _i}}}}}}]^T},\nonumber
\\&{{\bf{a}}_{r_i} }\buildrel \Delta \over= {[ {1,{e^{j {\Phi _{{r_i}}}}}, \ldots,{e^{j (N - 1){\Phi _{{r_i}}}}}} ]^T},\nonumber
\\&{{\bf{a}}_{v_i} } \buildrel \Delta \over= {[ {1,{e^{j {\Phi _{{v_i}}}}}, \ldots,{e^{j (M - 1){\Phi _{{v_i}}}}}} ]^T},\nonumber
\\
 &{\bf{a}}({\Phi _{{\theta _i}}},{\Phi _{{r _i}}},{\Phi _{{v_i}}})\buildrel \Delta \over={{\bf{a}}_{ \theta_i} }\otimes {{\bf{a}}_{r_i} }\otimes {{\bf{a}}_{v_i} }.
\end{align}
For brevity of expression, ${\bf a}_i$ denotes ${\bf{a}}({\Phi _{{\theta _i}}},{\Phi _{{r _i}}},{\Phi _{{v_i}}})$ in the later sections. 

For notational convenience, define the azimuth-parameter vector ${\boldsymbol{\theta }} \buildrel \Delta \over= {\rm{ }}{[{\theta _1},{\rm{ }}{\theta _2},...,{\theta _U}]^T} \in {\mathbb R^U}$, the range-parameter vector ${\bf{r}} \buildrel \Delta \over= {\rm{ }}{[{r_1},{\rm{ }}{r_2},...,{r_U}]^T} \in {\mathbb R^U}$, and the velocity-parameter vector ${\bf{v}} \buildrel \Delta \over= {\rm{ }}{[{v_1},{\rm{ }}{v_2},...,{v_U}]^T} \in {\mathbb R^U}$. By collecting the rotation vectors in~\eqref{eq:a123} of $U$ targets, the manifold matrix can be expressed as
\begin{align}\label{eq:AA123}
 {\bf{A}} \buildrel \Delta \over =\left[{\bf a}_1,{\bf a}_2,...,{\bf a}_U \right]\buildrel \Delta \over= {{\bf{A}}_{\boldsymbol \theta} } \odot {{\bf{A}}_{\bf r}} \odot {{\bf{A}}_{\bf v}},
\end{align}
where the manifold matrices ${{\bf{A}}_{\boldsymbol \theta} }$, ${{\bf{A}}_{\bf r}}$ and ${{\bf{A}}_{\bf v}}$ carrying the targets' azimuth, range and velocity information, respectively, are expressed as
\begin{align}\label{eq:A123}
&{{\bf{A}}_{\boldsymbol \theta} } \buildrel \Delta \over= \left[ {{{\bf{a}}_{ \theta_1} },{{\bf{a}}_{ \theta_2} }, \ldots,{{\bf{a}}_{ \theta_U} }} \right],\nonumber
\\&{{\bf{A}}_{\bf r}} \buildrel \Delta \over= \left[ {{{\bf{a}}_{r_1} },{{\bf{a}}_{r_2} }, \ldots,{{\bf{a}}_{ r_U} }} \right],\nonumber
\\&{{\bf{A}}_{\bf v}}\buildrel \Delta \over = \left[ {{{\bf{a}}_{r_1} },{{\bf{a}}_{ r_2} }, \ldots,{{\bf{a}}_{ r_U} }} \right].
\end{align}

By collecting signal observation in~\eqref{eq:cijk} for all antennas, subcarriers and symbols, the signal observation in column vector form ${\bf{z}} \in {{\mathbb C}^{LNM \times 1}}$ can be expressed as
\begin{align}\label{eq:signal observation}
{\bf{z}} = {\bf A \boldsymbol \alpha } + {\boldsymbol{\omega }},
\end{align}
where ${\bf{z}} \buildrel \Delta \over = [{{\bf{z}}_1};{{\bf{z}}_2};\ldots;{{\bf{z}}_L}]$ with ${\bf{z}}_l  = \big[{Z_{l,0}}[0],{Z_{l,1}}[0], \ldots,{Z_{l,M-1}}[0],\ldots,{Z_{l,0}}[N-1], \ldots, \\ {Z_{l,M-1}}[N-1] \big]^T$, the targets' backscatter coefficient vector $\boldsymbol{\alpha } \buildrel \Delta \over= {\rm{ }}{[{\alpha _1},{\rm{ }}{\alpha _2},...,{\rm{ }}{\alpha _U}]^T} \in \mathbb C^U$, and the term ${\boldsymbol{\omega }}$ is the noise vector stacked in the same manner as $\bf z$.

The objective of this paper is to utilize the obtained signal observation in~\eqref{eq:signal observation} to jointly estimate the targets' azimuth parameters $\theta_i$'s, range parameters $r_i$'s and velocity parameters $v_i$'s.

\section{Cram\'{e}r-Rao Bound Performance Analysis}\label{sec:CRB}

Before presenting the parameter estimation algorithms, this section derives the CRBs for the estimation of the azimuth, range and velocity parameters, which provides the lower bound on the squared RMSE of all unbiased estimation algorithms.

The unknown parameters to be estimated in~\eqref{eq:signal observation} can be stacked as the following column vector
\begin{align}\label{eq:Parameter}
&{\bf{\Gamma }} \buildrel \Delta \over= {\left[{{\boldsymbol{\theta }}};{{\bf{r}}};{{\bf{v}}};{\left( {{{\boldsymbol{\alpha }}^{\rm r}}} \right)};{\left( {{{\boldsymbol{\alpha }}^{\rm i}}} \right)}\right]} \in {\mathbb R^{5U}},
\end{align}
where the real-part vector and the imaginal-part vector of the targets' backscatter coefficient vector $\boldsymbol{\alpha }$ is denoted as ${{\boldsymbol{\alpha }}^{\rm r}} = {\mathop{\rm Re}\nolimits} \{ {\boldsymbol{\alpha }}\}\in {\mathbb R^U}$ and  ${{\boldsymbol{\alpha }}^{\rm i}} = {\mathop{\rm Im}\nolimits} \{ {\boldsymbol{\alpha }}\}\in {\mathbb R^U}$, respectively. The interest parameters to be estimated in the JARVE are ${{\bf{\Gamma }}_1} \buildrel \Delta \over= {\left[{{\boldsymbol{\theta }}}{;}{{\bf{r}}}{;}{{\bf{v}}}\right]} \in {{\mathbb {R}}^{3U}}$. However, the other unknown parameters ${{\bf{\Gamma }}_2} \buildrel \Delta \over={\left[{ {{{\boldsymbol{\alpha }}^{\rm r}}} };{{{{\boldsymbol{\alpha }}^{\rm i}}}}\right]} \in {\mathbb R^{2U}}$ still affect the CRBs on the estimation of ${{\bf{\Gamma }}_1}$.

The CRBs for JAVRE are given in the following theorem. 
{\textbf{\\\textit{Theorem 1}}}: From the signal observation in~\eqref{eq:signal observation}, the CRBs for JARVE of all targets are given by
\begin{align}\label{eq:CRBend}
&CRB({\bf{\Gamma }}_1) = \frac{{{\sigma ^2}}}{2}{\left[ { {\text{Re}}\{ ({{\bf{\Sigma }}^H}{{\bf{\dot A}}^H}{\bf{P}}_{\bf{A}}^ \bot{{\bf{\dot A}}}{\bf{\Sigma }})\} } \right]^{ - 1}},
\end{align}
where ${\bf{\Sigma }} = {{\bf{I}}_{3U}} \otimes {\text {{\text {diag}}}} ({\boldsymbol{\alpha }})$, the partial derivative of $\bf{A}$ with respect to ${{\bf{\Gamma }}_1}$ is ${\bf{\dot A}} = [{{\bf{\dot A}}_{\boldsymbol \theta} },{{\bf{\dot A}}_{\mathbf r}},{{\bf{\dot A}}_{\mathbf v}}]\in {{\mathbb C}^{LMN \times 3U}}$ with $ {{{\bf{\dot A}}}_{\boldsymbol \theta} } = {\partial {\bf{A}}}/{{\partial {\boldsymbol \theta} }},{{{\bf{\dot A}}}_{\mathbf r}} = {\partial{\bf{A}}}/{{\partial {\mathbf r}}},{{{\bf{\dot A}}}_{\mathbf v}} = {\partial{\bf{A}}}/{{\partial {\mathbf v}}}$, and ${\bf{P}}_{\bf{A}}^ \bot  = {\bf{I - A}}{\left( {{{\bf{A}}^H}{\bf{A}}} \right)^{ - 1}}{{\bf{A}}^H}$ denotes the orthogonal projection matrix of $\bf{A}$. 


\begin{proof}
The logarithmic likelihood function of ${\bf{z}}$ with respect to ${{\bf{\Gamma }}}$ can be derived as
\begin{align}\label{eq:lnLf}
\ell({\bf{z}}|{{\bf{\Gamma }}}) =LNM \ln{ \left(\pi \sigma ^2\right)} - \frac{{\parallel {\bf{z - A}\boldsymbol\alpha}\parallel _2^2}}{{{\sigma ^2}}}.
\end{align}


The CRBs of all unknown parameters ${{\bf{\Gamma }}}$ are as follows
\begin{align}\label{eq:CRB_all}
{{CRB}}({{\bf{\Gamma }}})  = {{\bf F}^{ - 1}},
\end{align}
where the Fisher information matrix $\bf F$ is 
\begin{align}\label{eq:Fisher matrix}
{\bf{F}} \buildrel \Delta \over= {\mathbb E}\left\{ {\frac{{\partial \ell ({\bf{z}}|{\bf{\Gamma}})}}{{\partial {\bf{\Gamma }}}}{{\left( {\frac{{\partial \ell ({\bf{z}}|{\bf{\Gamma}})}}{{\partial {\bf{\Gamma }}}}} \right)}^T}} \right\} = \left[ {\begin{array}{*{20}{c}}
{{{\bf{F}}_{11}}}&{{{\bf{F}}_{12}}}\\
{{\bf{F}}_{12}^T}&{{{\bf{F}}_{22}}}
\end{array}} \right].
\end{align}

According to~\eqref{eq:CRB_all} and~\eqref{eq:Fisher matrix}, the CRBs of the interested parameters can be expressed as
\begin{align}\label{eq:CRB_need}
{\rm{CRB}}({{\bf{\Gamma }}_1}) = {\left[ {{{\bf{F}}_{11}} - {{\bf{F}}_{12}}{\bf{F}}_{22}^{ - 1}{\bf{F}}_{12}^T} \right]^{ - 1}}.
\end{align}

To obtain the CRB, we derive the Fisher information matrix first. Each component of the partial derivative of logarithmic likelihood function in~\eqref{eq:lnLf} with respect to ${{\bf{\Gamma }}_1}$ is given by
\begin{align}\label{eq:diff Lnf1i}
{\left[ {\frac{{\partial \ell ({\bf{z}}|{\bf{\Gamma}})}}{{\partial {{\bf{\Gamma }}_1}}}} \right]_i} &=  \frac{2}{{{\sigma ^2}}}{\rm{Re}}\left\{ {{\alpha_u^H}\frac{{\partial {{\bf{a}}_u^H}}}{{\partial {{\left[ {{{\bf{\Gamma }}_1}} \right]}_i}}}{\boldsymbol{\omega}}} \right\},
\end{align}
for $i = 1,2,...,3U$, and $u=i\bmod U$.

Meanwhile, each component of the partial derivative of logarithmic likelihood function with respect to ${{\bf{\Gamma }}_2}$ is obtained as
\begin{align}\label{eq:diff Lnf2i}
{\left[ {\frac{{\partial \ell ({\bf{z}}|{\bf{\Gamma}})}}{{\partial {{\bf{\Gamma }}_2}}}} \right]_i}  &= \left\{ {\begin{array}{*{20}{c}}
{\frac{2}{{{\sigma ^2}}}{\rm{Re}}\left\{ {{{\bf{a}}_{i}^H}{\boldsymbol{\omega}}} \right\},}&{i = 1,...,U.}\\
{\frac{2}{{{\sigma ^2}}}{\rm{Im}}\left\{ {{{\bf{a}}_{i - U}^H}{\boldsymbol{\omega}}} \right\},}&{i = U+1,...,2U.}
\end{array}} \right.
\end{align}

Define the partial derivative of the manifold matrix ${\bf{A}}$ with respect to ${{\bf{\Gamma }}_1}$ as
\begin{align}\label{eq:diff A}
 {\bf{\dot A}} = \frac{{\partial {\bf{A}}}}{{\partial {{\bf{\Gamma }}_1}}} &= \left[ {\frac{{\partial {\bf{A}}}}{{\partial {{\left[ {{{\bf{\Gamma }}_1}} \right]}_1}}},\frac{{\partial {\bf{A}}}}{{\partial {{\left[ {{{\bf{\Gamma }}_1}} \right]}_2}}},...,\frac{{\partial {\bf{A}}}}{{\partial {{\left[ {{{\bf{\Gamma }}_1}} \right]}_{3U}}}}} \right] \in {{\mathbb C}^{LMN \times 3U}}.
\end{align}


From \eqref{eq:diff Lnf1i}, \eqref{eq:diff Lnf2i} and \eqref{eq:diff A}, the partial derivative of logarithmic likelihood function with respect to ${{\bf{\Gamma }}_1}$ and ${{\bf{\Gamma }}_2}$ are rewritten in terms of ${\bf{\dot A}}$ and $\bf A$, respectively, as follows 
\begin{align}\label{eq:diff Lnf}
\frac{{\partial \ell ({\bf{z}}|{\bf{\Gamma}})}}{{\partial {{\bf{\Gamma }}_1}}} &= \frac{2}{{{\sigma ^2}}}{\mathop{\rm Re}\nolimits} \left\{ {{{\bf{\Sigma }}^H}{{\bf{\dot A}}^H}{\boldsymbol{\omega }}} \right\}, \nonumber\\
\frac{{\partial \ell ({\bf{z}}|{\bf{\Gamma}})}}{{\partial {{\bf{\Gamma }}_2}}}& =\frac{2}{{{\sigma ^2}}}\left[ {{\rm{Re}}\left\{ {{{\bf{A}}^H}{\boldsymbol{\omega }}} \right\},{\rm{Im}}\left\{ {{{\bf{A}}^H}{\boldsymbol{\omega }}} \right\}} \right].
\end{align}

As derived in Appendix~\ref{app:F}, the block submatrices of Fisher information matrix in~\eqref{eq:Fisher matrix} are obtained as
\allowdisplaybreaks[2]
\begin{align}
{{\bf{F}}_{11}} &\buildrel \Delta \over= {\mathbb E}\left\{ {\frac{{\partial \ell ({\bf{z}}|{\bf{\Gamma}})}}{{\partial {{\bf{\Gamma }}_1}}}{{\left( {\frac{{\partial \ell ({\bf{z}}|{\bf{\Gamma}})}}{{\partial {{\bf{\Gamma }}_1}}}} \right)}^T}} \right\} \nonumber\\
 &= \frac{2}{{{\sigma ^2}}}{\mathop{\rm Re}\nolimits} {\{ {{\bf{\Sigma }}^H}{{\bf{\dot A}}^H}{\bf{\dot A\Sigma }}\} }\label{eq:F11},\\
{{\bf{F}}_{12}} &\buildrel \Delta \over= {\mathbb E}\left\{ {\frac{{\partial \ell ({\bf{z}}|{\bf{\Gamma}})}}{{\partial {{\bf{\Gamma }}_1}}}{{\left( {\frac{{\partial \ell ({\bf{z}}|{\bf{\Gamma}})}}{{\partial {{\bf{\Gamma }}_2}}}} \right)}^T}} \right\}\nonumber\\
 &= \frac{2}{{{\sigma ^2}}}{\left[ {{\mathop{\rm Re}\nolimits} \{ {{\bf{\Sigma }}^H}{{\bf{\dot A}}^H}{\bf{A}}\} , - {\mathop{\rm Im}\nolimits} \{ {{\bf{\Sigma }}^H}{{\bf{\dot A}}^H}{\bf{A}}\} } \right]}\label{eq:F12},\\
{{\bf{F}}_{22}} &\buildrel \Delta \over= {\mathbb E}\left\{ {\frac{{\partial \ell ({\bf{z}}|{\bf{\Gamma}})}}{{\partial {{\bf{\Gamma }}_2}}}{{\left( {\frac{{\partial \ell ({\bf{z}}|{\bf{\Gamma}})}}{{\partial {{\bf{\Gamma }}_2}}}} \right)}^T}} \right\}\nonumber\\
 &= \frac{2}{{{\sigma ^2}}}\left[ {\begin{array}{*{20}{c}}
{{\mathop{\rm Re}\nolimits} \left\{ {{{\bf{A}}^H}{\bf{A}}} \right\}}&{ - {\mathop{\rm Im}\nolimits} \left\{ {{{\bf{A}}^H}{\bf{A}}} \right\}}\\
{{\mathop{\rm Im}\nolimits} \left\{ {{{\bf{A}}^H}{\bf{A}}} \right\}}&{{\mathop{\rm Re}\nolimits} \left\{ {{{\bf{A}}^H}{\bf{A}}} \right\}}
\end{array}} \right]\label{eq:F22}.
\end{align}




Since ${{{\bf{A}}^H}{\bf{A}}}$ is nonsingular, the inverse of ${{\bf{F}}_{22}}$ in~\eqref{eq:F22} is obtained as 
\begin{align}\label{eq:F22inverse}
\setlength{\arraycolsep}{1.2pt}
{\bf{F}}_{22}^{ - 1} = \frac{2}{{{\sigma ^2}}}\left[ {\begin{array}{*{20}{c}}
{{\mathop{\rm Re}\nolimits} \left\{ {{{\left( {{{\bf{A}}^H}{\bf{A}}} \right)}^{ - 1}}} \right\}}&{ - {\mathop{\rm Im}\nolimits} \left\{ {{{\left( {{{\bf{A}}^H}{\bf{A}}} \right)}^{ - 1}}} \right\}}\\
{{\mathop{\rm Im}\nolimits} \left\{ {{{\left( {{{\bf{A}}^H}{\bf{A}}} \right)}^{ - 1}}} \right\}}&{{\mathop{\rm Re}\nolimits} \left\{ {{{\left( {{{\bf{A}}^H}{\bf{A}}} \right)}^{ - 1}}} \right\}}
\end{array}} \right],
\end{align}
which comes from the following fact (see APPENDIX E.6 of~\cite{CRB}) for any nonsingular matrix ${\bf{P}}$
\begin{align}\label{eq:result3}
\setlength{\arraycolsep}{1.2pt}
{\left[ {\begin{array}{*{20}{c}}
{{\mathop{\rm Re}\nolimits} \left\{ {\bf{P}} \right\}}&{ - {\mathop{\rm Im}\nolimits} \left\{ {\bf{P}} \right\}}\\
{{\mathop{\rm Im}\nolimits} \left\{ {\bf{P}} \right\}}&{{\mathop{\rm Re}\nolimits} \left\{ {\bf{P}} \right\}}
\end{array}} \right]^{ - 1}} \!=\! \left[ {\begin{array}{*{20}{c}}
{\mathop{\rm Re}\nolimits} \left\{ {{{\bf{P}}^{ - 1}}} \right\}&{ - {\mathop{\rm Im}\nolimits} \left\{ {{{\bf{P}}^{ - 1}}} \right\}}\\
{{\mathop{\rm Im}\nolimits} \left\{ {{{\bf{P}}^{ - 1}}} \right\}}&{\mathop{\rm Re}\nolimits} \left\{ {{{\bf{P}}^{ - 1}}} \right\}
\end{array}} \right].
\end{align}

From~\eqref{eq:result1} for complex matrix multiplication in Appendix~\ref{app:F}, substituting~\eqref{eq:F11},~\eqref{eq:F12},~\eqref{eq:F22} and~\eqref{eq:F22inverse} into~\eqref{eq:CRB_need} yields the CRBs in~\eqref{eq:CRBend} for the interested parameters, which completes the proof. 
\end{proof}

In other words, Theorem 1 states that the CRB for the $i$-th target is obtained as $CRB\left( {{\theta _i}} \right){\rm{ = }}{\left[ {CRB\left( {{\bf{\Gamma }}_1} \right)} \right]_{i,i}}$, $CRB\left( {{r_i}} \right) = {\left[ {CRB\left( {{\bf{\Gamma }}_1} \right)} \right]_{U + i,U + i}}$, and $CRB\left( {{v_i}} \right) = {\left[ {CRB\left( {{\bf{\Gamma }}_1} \right)} \right]_{2U + i,2U + i}}$, for $i = 1,2,...,U$. This implies that for the JARVE of multiple targets, the CRB performance for estimating each parameter of one target depends on the other parameters of all targets. In contrast, for the JARVE of single target, the CRB performance for estimating each parameter is independent of other parameters of the target, which is given in the following Corollary 1.
{\textbf{\\\textit{Corollary 1}}}:  The closed-form CRBs for JARVE of single target are given by
\begin{align}\label{eq:CRB123_s}
&CRB({\theta }){\text{ }}= \frac{{6}}{{{K_\theta }\gamma L\left( {{L^2} - 1} \right){d_\theta ^2 }}},\nonumber\\
&CRB({r}){\text{ }} = \frac{{6}}{{{K_r }\gamma N\left( {{N^2} - 1} \right){d_r^2}}},\nonumber\\
&CRB({v}){\text{ }}= \frac{{6}}{{{K_v }\gamma M\left( {{M^2} - 1} \right){d_v^2}}},
\end{align}
where $\gamma = |{\alpha }|^2/{{{\sigma ^2}}}$ denotes the SNR, 
${K_\theta } = NM,{K_r} = LM,{K_v} = LN $ are the corresponding number of snapshots in the antenna, subcarrier and symbol dimensions, respectively, the derivatives of $\Phi _\theta$, $\Phi_r$ and $\Phi _v$ are given by ${d_\theta } = 2\pi d\cos \theta /\lambda$, ${d_r}  = 4\pi \Delta f/c$, and ${d_v} = 4\pi {f_{\rm{c}}}\bar T/c$, respectively. 
\begin{proof}
From Theorem 1, Corollary 1 is obtained by proving the CRB matrix for the JARVE of single target is diagonal, and deriving the close-form expression for the diagonal elements. See details in Appendix~\ref{app:CRBs}.
\end{proof}

Clearly, Corollary 1 states that the CRB of each parameter is inversely proportional to the SNR, the snapshot number and the cubic of degree of freedom in the corresponding dimension. 


\section{Low-Complexity JARVE Algorithm Design}\label{sec:Proposed Algorithm}
In this section, the PI-2DMUSIC (i.e, Parallel Iterative-subspace-updating 2D-MUSIC) algorithm is proposed for low-complexity super-resolution JARVE. The proposed algorithm consists of the spatial smoothing step, the 3D-parameter initialization step, and the ISU 2D-MUSIC parallel estimation and 3D-parameter pairing step, which are presented the subsequent three subsections.  

\subsection{Spatial Smoothing}
The spatial smoothing operation is performed for the decoherence of signal observation components from multiple targets. Denote the new sub-signal observation sizes in space, frequency and time domains by $\tilde L,\tilde N,\tilde M$, respectively. Accordingly, the number of spatial smoothing sub-signal observations are ${S_{\rm a}} = L - \tilde L+1,{S_{\rm f}} = N - \tilde N+1$, and ${S_{\rm t}} = M - \tilde M+1$, respectively. The total number of sub-signal observations (i.e, snapshots) is $S={S_{\rm a}}{S_{\rm f}}{S_{\rm t}}$. And the $s$-th sub-signal observation, for $s={\tilde l}\tilde N\tilde M+{\tilde n}\tilde M+{\tilde m}$, can be expressed as 
\begin{align}\label{eq:C_smooth}
{{\bf{b}}}_{{\tilde l},{\tilde n},{\tilde m}} \buildrel \Delta \over = \left[{{\bf{ z}}_{{\tilde l},{\tilde m}}[{\tilde n}]};{{\bf{ z}}_{{\tilde l}+1,{\tilde m}}[{\tilde n}]}; \ldots;{{\bf{ z}}_{{\tilde l}+\tilde L-1,{\tilde m}}[{\tilde n}]}\right]\in {\mathbb C^{\tilde L\tilde N\tilde M}},
\end{align}
where ${{\bf{ z}}_{{\tilde l},{\tilde m}}[{\tilde n}]}=\big[ {Z_{\tilde l,\tilde m}}[\tilde n],{Z_{\tilde l,\tilde m +1}}[\tilde n], \ldots,{Z_{\tilde l,\tilde m+\tilde M-1}}[\tilde n],$\\ \ldots,${Z_{\tilde l,\tilde m}}[\tilde n + \tilde N-1], \ldots, {Z_{\tilde l,\tilde m + \tilde M-1}}[\tilde n + \tilde N-1]\big]^T\in {\mathbb C^{\tilde N\tilde M}}$ for the indices ${\tilde l}=0,1,...,{S_{\rm a}}-1,{\tilde n}=0,1,...,{S_{\rm f}}-1, {\tilde m}=0,1,...,{S_{\rm t}}-1$.


Furthermore, compared to~\eqref{eq:signal observation}, the $s$-th sub-signal observation ${{\bf{b}}}_{{\tilde l},{\tilde n},{\tilde m}}$ can be expressed as
\begin{align}\label{eq:sub-signal observations}
{{\bf{b}}_{{\tilde l},{\tilde n},{\tilde m}}} = {\tilde {\bf A}}{\bf{\Lambda }\boldsymbol \varpi}  + {\boldsymbol \upsilon}_{{\tilde l},{\tilde n},{\tilde m}}\in {\mathbb C^{\tilde L\tilde N\tilde M}},
\end{align}
where ${\tilde {\bf A}}\buildrel \Delta \over ={{\bf{A}}_{\boldsymbol \theta} }(1:\tilde L, :) \odot {{\bf{A}}_{\bf r}}(1:\tilde N, :) \odot {{\bf{A}}_{\bf v}}(1:\tilde M, :) \in {\mathbb C^{\tilde L\tilde N\tilde M \times U}}$ is the reconstructed manifold matrix, the diagonal matrix ${\bf{\Lambda }} = {\text {diag}}({\boldsymbol{\alpha }})$, the phase rotation vector of $s$-th sub-signal observation compared to the $0$-th sub-signal observation is $\boldsymbol \varpi  \buildrel \Delta \over = \left[ e^{j\left( {{\tilde l}{\Phi _{{\theta _1}}} + {\tilde n}{\Phi _{{r_1}}} + {\tilde m}{\Phi _{{v_1}}}} \right)},..., {e^{j\left( {{\tilde l}{\Phi _{{\theta _U}}} + {\tilde n}{\Phi _{{r_U}}} + {\tilde m}{\Phi _{{v_U}}}} \right)}}\right]^T$, and the term ${\boldsymbol \upsilon}_{{\tilde l},{\tilde n},{\tilde m}}$ is the spatially-smoothed noise vector reshaped in the same manner as ${\bf b}_{{\tilde l},{\tilde n},{\tilde m}}$. For brevity of expression, the ${\tilde {\bf{A}}_{\boldsymbol \theta} }$, ${\tilde {\bf{A}}_{\bf r} }$ and ${\tilde {\bf{A}}_{\bf v} }$ denote ${{\bf{A}}_{\boldsymbol \theta} }(1:\tilde L,:)$, ${{\bf{A}}_{\bf r} }(1:\tilde N, :)$ and ${{\bf{A}}_{\bf v} }(1:\tilde M,:)$, respectively in the latter sections.


To make the spatial-smoothing process clear, taking the slice of signal observation in any fixed frequency (i.e, the $n$-th subcarrier) for example, we demonstrate the smoothing operation for such signal observation in the space (i.e, antenna) domain and time (i.e, symbol) domain, as shown in Fig. 3 for $L=3, M=4, {S}_{\text{a}}={S}_{\text{t}}=2$. The smoothing operations in the space-frequency domains and the time-frequency domains are the same.
\begin{figure}[htbp]\label{fig3}
\centering
		\includegraphics[width=.89\columnwidth] {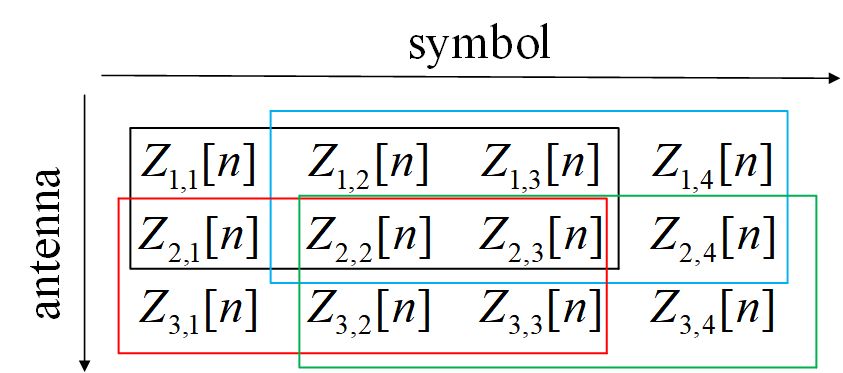}\\
\caption{Example of spatial smoothing for signal observations.}
\end{figure}

Then, the spatially-smoothed signal observation with $S$ snapshots is used to estimate the parameters $\boldsymbol \theta$, $\bf r$ and $\bf v$ in the following two subsections.

\subsection{3D-Parameter Initialization}\label{subsec:Proposed 1d}
For the joint estimation of multi-dimensional parameters, the spatial spectrum function is multi-variate, nonlinear and non-convex. Fortunately, the multi-dimensional spatial spectrum function has $U$ locally convex regions near to the peeks. In order to obtain initial points that are located to the $U$ locally convex regions, this subsection designs a low-complexity initialization method, which includes the proposed phase self-compensating ROOT-1D-MUSIC (PSCR-1DMUSIC) algorithm for initial 1D-parameter estimation, as well as the azimuth-range-velocity 3D-parameter pairing scheme.

Firstly, we take the range estimation process as an example to describe the PSCR-1DMUSIC algorithm, for ease of presentation.

From~\eqref{eq:cijk}, the phase rotation resulted by the range only varies with the frequency domain index $n$. The different signal observation components in the space and time domains can be exploited as different snapshots to improve the range estimation performance. The signal observation after frequency domain spatial smoothing has $S=LS_{\rm f}M$ snapshots, and is denoted by ${{\bf{b}}}_{{l},{\tilde n},{m}}\in {{\mathbb C}^{\tilde N }}$, for $l=0,1,...,L-1, \tilde n=0,1,...,S_{\rm f}-1,~m=0,1,...,M-1$ and $\tilde N=(N-S_{\rm f}+1)>U$. Thus, the sub-signal observation \eqref{eq:sub-signal observations} can be rewritten as
\begin{align}\label{eq:1D_ss}
{{\bf{b}}}_{{l},{\tilde n},{m}} &= {\tilde{\bf{A}}_{\bf r}}{\bf{\Lambda }}{\bf \Psi} {\boldsymbol \varpi }_{\rm r}   + {\boldsymbol \upsilon}_{{l},{\tilde n},{m}}\in {\mathbb C^{\tilde N}}\nonumber \\
 &=\sum\limits_{i = 1}^U \alpha _i{\tilde{\bf{ a}}_{r_i}}e^{j\left( {{l}{\Phi _{{\theta _i}}} + {\tilde n}{\Phi _{{r_i}}} + {m}{\Phi _{{v_i}}}} \right)} + {\boldsymbol \upsilon}_{{l},{\tilde n},{m}},
\end{align}
where the phase rotation matrix and vector relative to the ${{\bf{b}}_{{0},0,{0}}}$ are ${\bf \Psi} \buildrel \Delta \over = {\rm{diag}}\left\{ {{e^{j\left( {l{\Phi _{{\theta _1}}} + m{\Phi _{{v_1}}}} \right)}},...,{e^{j\left( {l{\Phi _{{\theta _U}}} + m{\Phi _{{v_U}}}} \right)}}} \right\}$ and ${\boldsymbol \varpi }_{\rm r} \buildrel \Delta \over = {\left[ {{e^{j\tilde n{\Phi _{{r_1}}}}},...,{e^{j\tilde n{\Phi _{{r_U}}}}}} \right]^T}$, respectively.

The covariance matrix of the sub-signal observation in \eqref{eq:1D_ss} is given by
\allowdisplaybreaks[3]
\begin{align}\label{eq:R1d}
&{\bf{R}}({{\bf r}}) = {\frac{1}{{L M}}}\sum\limits_{l = 1}^{ L} \sum\limits_{m = 1}^{ M}{\mathbb E}\left\{{{\bf{b}}}_{{l},{\tilde n},{m}}{{\bf{b}}}_{{l},{\tilde n},{m}}^H \right\} \nonumber\\
& \buildrel (a) \over=\sum\limits_{i = 1}^U |\alpha _i|^2{\tilde{\bf a}_{r_i} } {\tilde{\bf a}_{r_i}} ^H + {\sigma ^2}{{\bf{I}}_{\tilde N}}+{\frac{2}{L M}} \cdot \nonumber\\
&  \mathop{\rm Re} \left\{  {\sum\limits_{l,m = 1}^{L,M} {\sum\limits_{i,k \in {\calD_{ik}}}^{} \! \! {{\alpha _{i,k}}{e^{j \left(l{\Phi _{{\theta _{ik}}}} + m{\Phi _{{v_{ik}}}}\right)}}{\mathbb E}\left\{{e^{j\tilde n{\Phi _{{r_{ik}}}}}}\right\}{\tilde{\bf{a}}_{r_i} }\tilde{\bf{a}}_{r_k} ^H}}} \right\}\nonumber\\
& \buildrel (b) \over \approx {\tilde{\bf{A}}_{\bf r}}{\bf{\Lambda \Lambda}} ^H\tilde{\bf{A}}_{\bf r}^H + {\sigma ^2}{{\bf{I}}_{\tilde N}},
\end{align}
where the index set ${\calD_{ik}} = \{ (i,k)|1 \le i < k \le U\}$, ${\alpha _{i,k}}={\alpha _{i}}{\alpha _{k}^H}$, the phase rotation for different targets ${\Phi _{{\theta _{ik}}}} = {\Phi _{{\theta _i}}} - {\Phi _{{\theta _k}}},{\Phi _{{v_{ik}}}} = {\Phi _{{v_i}}} - {\Phi _{{v_k}}}$, ${\Phi _{{r_{ik}}}} = {\Phi _{{r_i}}} - {\Phi _{{r_k}}}$,  and (b) comes from the fact that the last summation term of (a) approximates 0 for typical values of $L$, $M$ and $S_{\rm f}$ in practice. 

In other words, the covariance-matrix summation in the space and time domains in \eqref{eq:R1d} can achieve self-compensation of the phases caused by all targets' azimuths and velocities, which results in that the last summation term of (a) is negligible. Thanks to this {\textit{phase self-compensation property}}, the range parameters can be individually estimated, being irrelevant to the estimation accuracy of other parameters.


By performing eigenvalue decomposition (EVD) to the covariance matrix in \eqref{eq:R1d}, the signal subspace and noise subspace of the signal observation can be obtained as
\begin{align}\label{eq:1dED}
\begin{array}{l}
{\bf{R}}({\bf{r}}) = {\bf{E\Omega  }}{{\bf{E}}^H}\\
 = \left[ {{{\bf{E}}_{\rm s}}~{\rm{ }}{{\bf{E}}_{\rm n}}} \right]\left[ {\begin{array}{*{20}{c}}
{{{\bf{\Omega }}_s}}&{{{\bf{0}}_{U \times \left( {\tilde N - U} \right)}}}\\
{{{\bf{0}}_{\left( {\tilde N - U} \right) \times U}}}&{{\sigma ^2}{{\bf{I}}_{\left( {\tilde N - U} \right)}}}
\end{array}} \right]\left[ {\begin{array}{*{20}{c}}
{{\bf{E}}_{\rm s}^H}\\
{{\bf{E}}_{\rm n}^H}
\end{array}} \right]\\
 = {{\bf{E}}_{\rm s}}{{\bf{\Omega }}_s}{\bf{E}}_{\rm s}^H + {\sigma ^2}{{\bf{E}}_{\rm n}}{\bf{E}}_{\rm n}^H,
\end{array}
\end{align}
where ${\bf{\Omega }}\in {{\mathbb R}^{\tilde N \times \tilde N}}$ is a diagonal matrix with non-increasing diagonal elements, ${{\bf{\Omega }}_s}\in {{\mathbb R}^{U \times U}}$ is a diagonal matrix with the largest $U$ eigenvalues of ${\bf{R}}({\bf{r}})$, the matrix ${{\bf{E}}_{\rm s}}\in {{\mathbb C}^{\tilde N \times U}}$ is the signal subspace, and the matrix ${{\bf{E}}_{\rm n}}\in {{\mathbb C}^{\tilde N \times (\tilde N-U)}}$ is the noise subspace.

By exploiting the orthogonality between the target's rotation vector $\tilde {\bf a}_{r_i}$'s and noise subspace $\bE_{\rm n}$, the range estimation is equivalent to find $U$ roots of the following $(2\tilde N-2)$-order polynomial (i.e, 1D-root function) 
\begin{align}\label{eq:g1d1}
{g_{\rm pr}}({\kappa_r})=\kappa_r^{\tilde N - 1}{{\tilde {\bf{a}}}^H_{r}}{({\kappa_r})}{{\bf{E}}_{\rm n}}{{\bf{E}}_{\rm n}^H}{{\tilde {\bf{a}}}_{r}}({\kappa_r}),
\end{align}
where the variable ${\kappa_r} \buildrel \Delta \over = {e^{-j2\pi 2r\Delta f/c}}$. 

To avoid high computational complexity caused by spatial spectrum calculation, the Pisarenko decomposition~\cite{root-music} can be performed for ${g_{\rm pr}}({\kappa_r})$ in \eqref{eq:g1d1} to find the roots $\hat{\kappa}_{r_j}$, for $j=1,2,..,U$, and obtain $\hat r_j$ as follows
\begin{align}\label{eq:3Dr}
{\hat r_j}=  - \frac{\angle {\kappa_{r_j}}}{4\pi \Delta f}.
\end{align}

Similarly, let ${\kappa_\theta} \buildrel \Delta \over = {e^{j2\pi d \sin \theta /\lambda}}$ and ${\kappa_v} \buildrel \Delta \over = {e^{j2\pi 2{v}{f_{\rm{c}}}\bar T/c}}$, the targets' azimuths and velocities can be estimated as
\begin{align}\label{eq:3D123}
{\hat \theta_i} &=  \frac{180}{\pi} {\text{arcsin}} \left[ \frac{{\lambda \angle{\kappa_{{\theta_i}}}}}{{2\pi d}} \right], \; \text{for}  \; i=1,2,...,U,\nonumber\\
{\hat v_k} &= \frac{{\angle{\kappa_{{v_k}}}c}}{{4\pi {f_{\rm{c}}}\bar T}}, \; \text{for} \; k=1,2,...,U.
\end{align}

Secondly, we propose an efficient pairing scheme for all ${\hat {\theta_i}}$'s, $\hat r_j$'s and $\hat v_k$'s, via performing the maximum likelihood estimation (MLE) for the target backscatter coefficients $\boldsymbol \alpha$ and following minimum mean square error (MMSE) criterion for signal observation $\bf z$ in \eqref{eq:signal observation}.

Clearly, there are $(U!)^2$ possible combinations for matching the estimated parameters ${\hat \theta _{i_q}},{\hat r_{j_q}}$ and ${\hat v_{k_q}}$, for $i_q, j_q, k_q = 1,2,...,U$, into the $U$ targets. The correspondingly reconstructed manifold matrices are denoted by ${{\bf{\hat A}}_q}\left({\hat \theta _{i_q}},{\hat r_{j_q}},{\hat v_{k_q}}\right) \in {\mathbb C^{LMN \times U}}, \; \text{for} \; q = 1,2,...,{(U!)^2}$. From \eqref{eq:signal observation}, the $(U!)^2$ MLEs of the target backscatter coefficients $\boldsymbol \alpha$ are obtained by traversing each ${{\bf{\hat A}}_q}$ as
\begin{align}\label{eq:JML1}
{{{\hat {\boldsymbol{\alpha}} }}_q} = {\bf{\hat A}}_q^\dag {\bf{z}} = {({\bf{\hat A}}_q^H{{\bf{\hat A}}_q})^{ - 1}}{\bf{\hat A}}_q^H{\bf{z}}.
\end{align}

The optimal pairing combination $({\hat \theta_{i_{q^{\star}}}},{\hat r_{j_{q^{\star}}}},{\hat v_{k_{q^{\star}}}})$ denoted by $({\hat \theta_{i}^0},{\hat r_{i}^0},{\hat v_{i}^0})$ can be obtained via the following MMSE criterion for $\bf z$
\begin{align}\label{eq:JML2}
{\hat {\boldsymbol{\alpha}}}_{q^\star} = \mathop {\arg \min  }\limits_{{{\hat \alpha }_q}} ||{\bf{z}} - {\bf{\hat A}}_q^H{{{\hat {\boldsymbol{\alpha}} }}_q}|{|^2}, \;  \forall  q.
\end{align}

So far, the 3D-parameter initialization is completed. The steps is summarized in the following Algorithm 1.
\begin{algorithm}[!ht]
    \caption{3D-parameter Initialization}
    \label{alg:3D-parameter Initialization}
    \renewcommand{\algorithmicrequire}{\textbf{Input:}}
    \renewcommand{\algorithmicensure}{\textbf{Output:}}
    \begin{algorithmic}[1]
        \REQUIRE signal observation $\bf z$, number of targets $U$  
		\STATE Perform spatial smoothing to obtain ${\bf{b}}_{{\rm r} ,s}$ as~\eqref{eq:1D_ss}.
		\STATE Calculate covariance matrix ${\bf{R}}({\bf{r}})$ as~\eqref{eq:R1d}. 
        \STATE Perform EVD to obtain
  noise subspace ${{\bf{E}}_{\rm n}}{\kern 1pt}$ as~\eqref{eq:1dED}.
		\STATE Perform Pisarenko decomposition for ${g_{\rm pr}}({\kappa_r})$ in~\eqref{eq:g1d1} to obtain ${\hat r_j}$ for $j = 1,2, \ldots,U$ as \eqref{eq:3Dr}.
		\STATE Replace ${r}$ with $\theta$ and ${v}$ and repeat steps 1-4 to obtain ${\hat \theta_i},{\hat v_k}$ for $i,k = 1,2, \ldots,U$.
		\STATE Perform MLE for $\boldsymbol \alpha$ as~\eqref{eq:JML1} to obtained ${{{\hat {\boldsymbol{\alpha}} }}_q}$, for $q = 1,2,...,(U!)^2$.
		\STATE Pair the estimated parameters ${\hat \theta _{i_q}},{\hat r_{j_q}}$ and ${\hat v_{k_q}}$ via the MMSE criterion as~\eqref{eq:JML2}, and obtain the optimal pairing $({\hat \theta_{i}^0},{\hat r_{i}^0},{\hat v_{i}^0})$ for $i = 1,2, \ldots,U$.
		\ENSURE optimal pairing $({\hat \theta_{i}^0},{\hat r_{i}^0},{\hat v_{i}^0})$'s.  
    \end{algorithmic}
\end{algorithm}

Let $\tilde Z_1>\tilde Z_2>\tilde Z_3$ is a decreasing sort of the spatially-smoothed 3D sub-signal observation sizes $\tilde L$, $\tilde N$ and $\tilde M$. The correspondingly original signal observation sizes are $Z_1$, $Z_2$ and $Z_3$, respectively. Taking into account the initial 3D-parameter estimation and pairing method, the main computational complexity of Algorithm 1 comes from step 2, step 3 and step 4, which is ${\rm O}\left( {{{ {{{\tilde Z}_1^2}}}}{S_1}} \right)$, ${\rm O}\left( {{{ {{{\tilde Z}_1^3}} }}} \right)$ and ${\rm O}\left( {{{ {{{\tilde Z}_1^3}}}}} \right)$, respectively, where the number of snapshots is $S_1=(Z_1-\tilde Z_1+1)Z_2Z_3$. Thus, Algorithm 1 only occupies very low complexity in the whole PI-2DMUSIC algorithm.
\subsection{ Iterative-subspace-updating 2D-MUSIC Parallel Estimation and 3D-Parameter Pairing}\label{subsec: Proposed 2D}

In this section, after slicing the 3D-signal observations in each dimension and performing spatial smoothing, the resulting three 2D-signal observations are first used to estimate the 2D parameters by the proposed iterative-subspace-updating 2D-MUSIC (ISU-2DMUSIC) algorithm in a parallel manner, and the estimated 2D parameters are matched to the 3D-parameter pairs for all targets by following the minimum-distance criterion.


For brevity of presentation, we first take the joint azimuth and velocity estimation as an example to describe the proposed ISU-2DMUSIC algorithm in the sequel.

The 2D-signal observation in the space-time domains is obtained by slicing the 3D-signal observations in~\eqref{eq:sub-signal observations} from the frequency domain. The spatially-smoothed 2D-signal observations with $S=S_{\rm a}NS_{\rm t}$ snapshots are denoted by ${{\bf{b}}}_{{\tilde l},{n},{\tilde m}}\in {{\mathbb C}^{\tilde L \tilde M}}$ for $l=0,1,...,S_{\rm a}-1, \tilde n=0,1,...,N-1,~m=0,1,...,S_{\rm t}-1$ and $\tilde L\tilde M>U$. Thus, the sub-signal observation \eqref{eq:sub-signal observations} can be rewritten as
\begin{align}\label{eq:2D_ss}
&{\bf{b}}_{\tilde l,n,\tilde m} = {\tilde {\bf A}}({\boldsymbol \theta},{\bf v}) {\bf{\Lambda }}{\bf \Psi}_{\rm r} {\boldsymbol \varpi }   + {\boldsymbol \upsilon}_{{\tilde l},{n},{\tilde m}}\in {\mathbb C^{{\tilde L}{\tilde M}}}\nonumber\\
&=\sum\limits_{i = 1}^U \alpha _i{\tilde{\bf{ a}}_{\theta_i,v_i}}e^{j\left( {{\tilde l}{\Phi _{{\theta _i}}} + {n}{\Phi _{{r_i}}} + {\tilde m}{\Phi _{{v_i}}}} \right)} + {\boldsymbol \upsilon}_{{\tilde l},{n},{\tilde m}},
\end{align}
where the 2D manifold matrix is ${\tilde {\bf A}}({\boldsymbol \theta},{\bf v})\buildrel \Delta \over={\tilde{\bf{A}}_{\boldsymbol \theta}}\odot{\tilde{\bf{A}}_{\bf v}}$, the 2D rotation vector is ${\tilde{\bf{ a}}_{\theta_i,v_i}}\buildrel \Delta \over={\tilde{\bf{ a}}_{\theta_i}}\otimes{\tilde{\bf{a}}_{v_i}}$, the phase rotation matrix and vector relative to the ${{\bf{b}}_{{0},0,{0}}}$ are ${\bf \Psi}_{\rm r}  \buildrel \Delta \over = {\rm{diag}}\left\{ {{e^{jn{\Phi _{{r_1}}}}},...,{e^{jn{\Phi _{{r_U}}}}}} \right\}$ and ${\boldsymbol \varpi }\buildrel \Delta \over = {\left[ {{e^{j\left( {\tilde l{\Phi _{{\theta _1}}} + \tilde m{\Phi _{{v_1}}}} \right)}},...,{e^{j\left( {\tilde l{\Phi _{{\theta _U}}} + \tilde m{\Phi _{{v_U}}}} \right)}}} \right]^T}$, respectively.

The covariance matrix of the sub-signal observation in \eqref{eq:2D_ss} can be computed as 
\allowdisplaybreaks[3]
{\fontsize{9.5}{10}
\begin{align}\label{eq:R2d}
&{\bf{R}}({{\boldsymbol \theta,\bf v}}) = \frac{1}{N}\sum\limits_{n = 1}^{ N}{\mathbb E}\left\{{{\bf{b}}}_{{\tilde l},{n},{\tilde m}}{{\bf{b}}}_{{\tilde l},{ n},{\tilde m}}^H \right\} \nonumber\\
& \buildrel (a) \over=\sum\limits_{i = 1}^U |\alpha _i|^2{\tilde{\bf{ a}}_{\theta_i,v_i}} {\tilde{\bf{ a}}_{\theta_i,v_i}}^H + {\sigma ^2}{{\bf{I}}_{\tilde L\tilde M}}+{\frac{2}{{N}}}\cdot \nonumber\\
&{\mathop{\rm Re}\nolimits}\Bigg\{\!{\sum\limits_{n = 1}^{N}\!{\sum\limits_{i,k \in {\calD_{ik}}}}\!\! {{\alpha _{i,k}}\!{e^{jn{\Phi _{r_{ik}}}}}{\mathbb E}\left\{{e^{j\left(\tilde l{\Phi _{{\theta _{ik}}}} + \tilde m{\Phi _{{v_{ik}}}}\right)}}\right\}{\tilde{\bf{ a}}_{\theta_i,v_k}}\tilde{\bf{ a}}_{\theta_i,v_k}^H}}\!  \Bigg\}\nonumber\\
& \buildrel (b) \over \approx {\tilde {\bf A}({\boldsymbol \theta},{\bf v})}{\bf{\Lambda \Lambda}} ^H{{\tilde {\bf A}}^H({\boldsymbol \theta},{\bf v})} + {\sigma ^2}{{\bf{I}}_{\tilde L\tilde M}},
\end{align}}
where (b) is due to the fact that the last summation term of (a) approximates 0 in practice.




Following the similar EVD in \eqref{eq:1dED}, the noise subspace for the covariance matrix $\bf{R}({\boldsymbol{\theta }},{\bf{v}})$ in \eqref{eq:R2d} can be obtained and denoted by ${\bf E}_n\in {{\mathbb C}^{\tilde L \tilde M\times (\tilde L \tilde M-U)}}$. From the orthogonality between the target's rotation vector $\tilde {\bf a}_{\theta_i,v_i}$'s  and noise subspace ${\bf E}_n$, the azimuth and velocity parameters can be jointly estimated by minimizing the following two-variable 
 least square function 
\begin{align}\label{eq:object_function}
{{G}}({{\boldsymbol{\kappa}}}) = {{{\tilde{\bf a}}}}^H{({{\boldsymbol{\kappa}}})}{{\bf{E}}_{\rm n}}{{\bf{E}}_{\rm n}^H}{\tilde{ \bf a}}({{\boldsymbol{\kappa}}}),
\end{align}
where the 2D phase rotation vector ${\tilde{ \bf a}}({{\boldsymbol{\kappa}}})\buildrel \Delta \over={\tilde{\bf{ a}}_{\theta_i}}\otimes{\tilde{\bf{a}}_{v_i}}$ for  $\begin{array}{l}{{\boldsymbol{\kappa}}} \buildrel \Delta \over = {[{\kappa_\theta },{\kappa_v}]^T}
 = {[{e^{j2\pi d\sin \theta /\lambda }},{e^{j2\pi {{2v{f_{\rm{c}}}\bar T} \mathord{\left/
 {\vphantom {{2v{f_{\rm{c}}}\bar T} c}} \right.
 \kern-\nulldelimiterspace} c}}}]^T}.\end{array}$

To avoid the extremely high complexity of exhaustive search in conventional 2D-MUSIC algorithm, the Levenberg-Marquardt (LM) method is utilized to optimize the spatial spectrum, which is briefly described herein.

The objective function in \eqref{eq:object_function} can be rewritten as
\begin{align}\label{eq:3D7}
G({{\boldsymbol{\kappa}}}) = \frac{1}{2}{{\bf{g}}^H}({{\boldsymbol{\kappa}}}){\bf{g}}({{\boldsymbol{\kappa}}}),
\end{align}
where the vector-valued function ${\bf{g}}({{\boldsymbol{\kappa}}})$ as
\begin{align}\label{eq:gz}
{\bf{g}}({{\boldsymbol{\kappa}}}) = {\sqrt 2 {{\bf{E}}_{\rm n}^H}{\tilde {\bf a}}({{\boldsymbol{\kappa}}})}\in{\mathbb C^{\tilde L\tilde N\tilde M}}.
\end{align}

The first-order Peano-type Taylor expansion of ${\bf{g}}({{\boldsymbol{\kappa}}})$ is 
\begin{align}\label{eq:3D8}
{\bf{g}}({{\boldsymbol{\kappa}}} + {\bf{h}}) = {\bf{g}}({{\boldsymbol{\kappa}}}) + {\bf{J}}({{\boldsymbol{\kappa}}}){\bf{h}} + {\rm O}({{\bf{h}}^H}{\bf{h}}),
\end{align}
where the Jacobian matrix ${\bf{J}}({{\boldsymbol{\kappa}}})$ of ${\bf{g}}({{\boldsymbol{\kappa}}})$ is given by

\begin{align}\label{eq:Jacobian}
{\bf{J}}({{\boldsymbol{\kappa}}}) = \sqrt 2{\bf E}_{\rm n}^H\left[ {\frac{{\partial {\tilde {\bf a}}\left( {{\boldsymbol{\kappa}}} \right)}}{{\partial {\kappa_\theta }}},\frac{{\partial {\tilde {\bf a}}\left( {{\boldsymbol{\kappa}}} \right)}}{{\partial {\kappa_v}}}} \right].
\end{align}

By substituting~\eqref{eq:3D8} into~\eqref{eq:3D7}, the second-order approximation of ${\bf{G}}({{\boldsymbol{\kappa}}})$ can be obtained as 
\begin{align}\label{eq:3D9}
G({{\boldsymbol{\kappa}}} + {\bf{h}}) \approx \frac{1}{2}{{\bf{g}}^H}{\bf{g}} + {{\bf{h}}^H}{\bf{J}}^H {\bf{g}} + \frac{1}{2}{{\bf{h}}^H}{{\bf{J}}^H}{\bf{Jh}}.
\end{align}

By introducing the damping term $\frac{1}{2}\mu {{\bf{h}}^H}{\bf{h}}$ to \eqref{eq:3D9} to keep the Hessian matrix positive definite~\cite{Levenberg,Marquardt}, the optimal iteration step size is given by
\begin{align}\label{eq:3D10}
{{\bf{h}}_{\rm lm}} &= \mathop {\arg \min }\limits_{\bf{h}} \frac{1}{2}{{\bf{g}}^H}{\bf{g}} + {{\bf{h}}^H}{{\bf{J}}^H}{\bf{g}} + \frac{1}{2}{{\bf{h}}^H}{{\bf{J}}^H}{\bf{Jh}} + \frac{1}{2}\mu {{\bf{h}}^H}{\bf{h}},\nonumber\\
&=  - {({{\bf{H}}_{\rm es}} + \mu {\bf{I}})^{ - 1}}{{\bf{J}}^H}{\bf{g}},
\end{align}
where the Hessian matrix ${{\bf{H}}_{\rm es}} = {{\bf{J}}^H}{\bf{J}}$, the damping coefficient $\mu  > 0$, and the positive-definite modified-Hessian matrix is $({{\bf{H}}_{\rm es}} + \mu {\bf{I}})$. 

Following~\cite{Levenberg,Marquardt}, let $\tau$ be a predefined small value, the initial Hessian matrix ${\bf{H}}_{\rm es}^0$ and the initial damping coefficient $\mu$ are determined as 
\begin{align}\label{eq:3D12}
{\bf{H}}_{\rm es}^0 &= {\bf{J}}{({{{\boldsymbol{\kappa}}}_0})^H}{\bf{J}}({{{\boldsymbol{\kappa}}}_0}),\\
\mu &= \tau \max \left\{ {\text {diag}}\{ {\bf{H}}_{\rm es}^0\}\right\},
\end{align}
and the gain ratio for updating $\mu$ is given by 
\begin{align}\label{eq:cost}
\rho  = \frac{{G({{\boldsymbol{\kappa}}}) - G({{\boldsymbol{\kappa}}} + {{\bf{h}}_{\rm lm}})}}{{L(0) - L({{\bf{h}}_{\rm lm}})}}.
\end{align}


The optimal ${{{\boldsymbol{\kappa}}}^ \star  }$ can be obtained by the LM method, see details in the steps 5-20 in the latter Algorithm 2.
Using the $U$ initial points ${\hat{  \boldsymbol{\kappa}}}_i^0=\left[{\hat \kappa _{\theta _i^0}} ,{\hat \kappa _{v_i^0}}\right]^T$ from Algorithm 1 located in $U$ locally convex regions of 2D-spatial-spectrum function, the LM algorithm are performed by $U$-times to estimate all targets' 2D parameters. 

In order to further improve the 2D-parameter estimation performance, we propose the iterative-subspace-updating (ISU) processing before each use of the LM method to eliminate the interference of echo signals from the estimated target.

To facilitate interference elimination, the targets with stronger echo (i.e, lower minimum value of root function) are preferentially estimated. An average 1D-root function ${{\rm{\bar g}}_{\rm pr}} ({\hat{\boldsymbol{\kappa}}}_i^0)$ is defined as the weighted sum of the 1D-root functions in \eqref{eq:g1d1} from two dimensions, i.e,
\begin{align}\label{eq:ISU0}
{{\rm{\bar g}}_{\rm pr}}({\hat{\boldsymbol{\kappa}}}_i^0) =  \frac{{{{\rm{g}}_{\rm pr}}({{{\hat \kappa}}}_{\theta_i^0})}}{{{{\tilde L}^2}}} + \frac{{{{\rm{g}}_{\rm pr}}({{\hat{\kappa}}}_{v_i^0})}}{{{{\tilde M}^2}}},
\end{align}
where the weight is the reciprocal of the square for the corresponding length of rotation vector ${\tilde{\bf{ a}}_{\theta_i}}$ and ${\tilde{\bf{a}}_{v_i}}$.


By using ${{\rm{\bar g}}_{\rm pr}} ({\hat{\boldsymbol{\kappa}}}_i^0)$ in \eqref{eq:ISU0}, the initial iteration points ${\hat{  \boldsymbol{\kappa}}}_i^0$'s can be sorted to obtain ${{\boldsymbol{\kappa}}}_i^\prime$'s as
\begin{align}\label{eq:ISU1}
&{{\rm{\bar g}}_{\rm pr}}({{\boldsymbol{\kappa}}}_1^\prime) \le {{\rm{\bar g}}_{\rm pr}}({{\boldsymbol{\kappa}}}_2^\prime) \le  \ldots  \le {{\rm{\bar g}}_{\rm pr}}({{\boldsymbol{\kappa}}}_U^\prime),\\
&\; \text{for} \; \boldsymbol{\kappa}_i^\prime \in \big\{{\hat{  \boldsymbol{\kappa}}}_i^0, 
\; \text{for} \; i=1,2,\ldots,U \big\}.
\end{align}

Let ${{\bf{E}}_{\rm n}^1={\bf{E}}_{\rm n}}$ denote the noise subspace before the first use of LM method. The optimal ${{{\boldsymbol{\kappa}}}^\star_i}$ can be obtained by the $i$-th LM method for $i=1,...,U$, see details in the steps 5-20. After the $(i-1)$-th use of LM method, the estimated rotation vector ${\tilde{\bf a}}\left({{\boldsymbol{\kappa}}}_{i-1}^\star  \right)$ is projected into the orthogonal complementary space of the noise subspace ${\bf{E}}_{\rm n}^{i-1}$ to update the noise subspace ${\bf{E}}_{\rm n}^{i}$ as follow
\begin{align}\label{eq:ISU3}
{\bf{E}}_{\rm n}^{i} &= [{\bf{E}}_{\rm n}^{i-1},{\bf{v}}_{\rm norm}^{i}], \nonumber\\
{\bf{v}}_{\rm norm}^{i}  &= \frac{{\left[ {{\bf{I}} - {{\bf{E}}_{\rm n}^{i-1}}{(\bf{E}}_{\rm n}^{i-1})^H} \right]{\tilde{\bf a}}\left({{\boldsymbol{\kappa}}}_{i-1}^\star  \right)}}{{\left\| {\left[ {{\bf{I}} - {{\bf{E}}_{\rm n}^{i-1}}{(\bf{E}}_{\rm n}^{i-1})^H} \right]{\tilde{\bf a}}\left({{\boldsymbol{\kappa}}}_{i-1}^\star \right)} \right\|}_2}.
\end{align}

The objective vector-valued function in~\eqref{eq:gz} and Jacobian matrix in~\eqref{eq:Jacobian} are updated, respectively, as follows 
\begin{align}
{\bf{g}}_i&= {\sqrt 2\left({{\bf{E}}_{\rm n}^i}\right)^H{\tilde{\bf a}}({{\boldsymbol{\kappa}}})},\label{eq:gi}\\
{\bf{J}}_i &= \sqrt 2\left({{\bf{E}}_{\rm n}^i}\right)^H\left[ {\frac{{\partial {\tilde {\bf a}}\left( {{\boldsymbol{\kappa}}} \right)}}{{\partial {\kappa_\theta }}},\frac{{\partial {\tilde {\bf a}}\left( {{\boldsymbol{\kappa}}} \right)}}{{\partial {\kappa_v}}}} \right].\label{eq:Ji}
\end{align}

The steps of ISU-2DMUSIC algorithm are summarized in the following Algorithm 2.
\begin{algorithm}[!ht]\label{Algorithm2}
    \caption{ISU-2DMUSIC Algorithm}
    \label{alg:ISU-2DMUSIC Algorithm}
    \renewcommand{\algorithmicrequire}{\textbf{Input:}}
    \renewcommand{\algorithmicensure}{\textbf{Output:}}
    \begin{algorithmic}[1]
       \REQUIRE  signal observation $\bf z$, $U$ initial points ${{\boldsymbol{\kappa}}}_i^0$'s, maximum iteration number $q _{\rm max}$, function tolerance ${\varepsilon _1}$, step size tolerance ${\varepsilon _2}$ and initial coefficient $\tau$ 
       	\STATE Calculate covariance matrix ${\bf{R}}({\boldsymbol{\theta }},{\bf{v}})$ as~\eqref{eq:R2d}.
		\STATE Perform EVD ${\bf{R}}({\boldsymbol{\theta }},{\bf{v}})$ to obtain ${{\bf{E}}_{\rm n}}{\kern 1pt}$ as~\eqref{eq:1dED}.
		\STATE Obtain ${{\boldsymbol{\kappa}}}_i^\prime$ by sorting ${\hat{  \boldsymbol{\kappa}}}_i^0$ for $ i = 1,2, \ldots,U$ as~\eqref{eq:ISU1}.
		\FOR{$i = 1,2, \ldots,U$}
		\STATE Update ${\bf{E}}_{\rm n}^{i}$ as~\eqref{eq:ISU3}.
		\STATE Update ${\bf{g}}_i$ as~\eqref{eq:gi}, ${\bf{J}}_i$ as~\eqref{eq:Ji} and ${{\boldsymbol{\kappa}}_0}={{\boldsymbol{\kappa}}_i^\prime}$.
       \STATE Initialize $q: = 0;\nu : = 2;{{\boldsymbol{\kappa}}}: = {{{\boldsymbol{\kappa}}}_0};{{\bf{H}}_{\rm es}}: = {\bf{J}}_i^H{\bf{J}}_i$, $found: = \left( {{{\left\| {{{\bf{J}}_i^H}{\bf{g}}_i} \right\|}_\infty } \le {\varepsilon _1}} \right);\mu : = \tau \max \left\{ {\text {diag}}\{ {\bf{H}}_{\rm es}^0\}\right\}$.
		\STATE ${\bf{while}}(not~found) {\bf{and}} \left( {q < q _{\rm max}} \right)$
		\STATE ~~$q: = q + 1;$ obtain ${{\bf{h}}_{\rm lm}}$ as~\eqref{eq:3D10}
		\STATE ~~${\bf{if}}\left\| {{{\bf{h}}_{\rm lm}}} \right\| \le {\varepsilon _2}\left( {\left\| {{\boldsymbol{\kappa}}} \right\| + {\varepsilon _2}} \right)$
		\STATE ~~~~$found: = true$
		\STATE ~~${\bf{else}}$
		\STATE ~~~~${{{\boldsymbol{\kappa}}}_{\rm new}}: = {{\boldsymbol{\kappa}}} + {{\bf{h}}_{\rm lm}}$, update $\rho$ as~\eqref{eq:cost}.   
		\STATE ~~~~${\bf{if}}\rho  > 0$
		\STATE ~~~~~~${{\boldsymbol{\kappa}}}: = {{{\boldsymbol{\kappa}}}_{\rm new}};{{\bf{H}}_{\rm es}}: = {\bf{J}}_i^H{\bf{J}}_i$
		\STATE ~~~~~~$found: = \left( {{{\left\|  {{{\bf{J}}_i^H}{\bf{g}}_i} \right\|}_\infty } \le {\varepsilon _1}} \right)$
		\STATE ~~~~~~$\mu : = \mu \max \left\{ {1/3,1 - {{(2\rho  - 1)}^3}} \right\};\nu : = 2$
		\STATE ~~~~${\bf{else}}$
		\STATE ~~~~~~$\mu : = \mu \nu ;\nu : = 2\nu $
		\STATE{\bf{end}}

		\STATE Using optimal iteration output ${{\boldsymbol{\kappa}}_i^ \star}={{\boldsymbol{\kappa}}}$ to obtain $({\hat \theta _i},{\hat v_i})$ accroding to~\eqref{eq:3D123}.  
		\ENDFOR
		\ENSURE $({\hat \theta _i},{\hat v_i})$ for $i=1,2,...,U$.  
    \end{algorithmic}
\end{algorithm}

It is noticed that the joint azimuth-range estimation and the joint range-velocity estimation can be achieved by following similar process of the above joint azimuth-velocity estimation, which is omitted due to limited space.

By performing the proposed ISU-2DMUSIC algorithms for three times, the three sequences of 2D-parameter pairs $(\hat \theta_i, \hat v_i)$, $(\hat \theta_j, \hat r_j)$ and $(\hat r_k, \hat v_k)$, for $i,j,k=1,2,...,U$, can be obtained. Following the minimum-distance criterion, these pairs are re-matched and yield the 3D-parameter estimation as follows
\begin{align}\label{eq:pair}
&\hat \theta ^\star_i =\frac{\hat \theta_{i}+\hat \theta_{j_i}}{2}, \; \text{for} \; j_i=\mathop {\arg \min }\limits_{j}\left\|\hat \theta_{i}-\hat \theta_{j}\right\|_2,\nonumber\\
&\hat r ^\star_i =\frac{\hat r_{j_i}+\hat r_{k_{j_i}}}{2}, 
\hat v^\star_i =\frac{\hat v_{i}+\hat v_{k_{j_i}}}{2}, \nonumber\\
\; &\text{for} \; k_{j_i}=\mathop {\arg \min }\limits_{k}\left\|\hat r_{j_i}-\hat r_{k_{j_i}}\right\|_2+\left\|\hat v_{i}-\hat v_{k_{j_i}}\right\|_2.
\end{align}

Finally, combining the Algorithm 1 and Algorithm 2, the steps of the proposed PI-2DMUSIC algorithm is summarized in the following Algorithm 3.
\begin{algorithm}[!ht]
    \caption{PI-2DMUSIC}
    \label{alg:PI-2DMUSIC}
    \renewcommand{\algorithmicrequire}{\textbf{Input:}}
    \renewcommand{\algorithmicensure}{\textbf{Output:}}
    \begin{algorithmic}[1]
        \REQUIRE  signal observation $\bf z$, number of targets $U$  
		\STATE Perform spatial
smoothing to obtain  ${\bf{b}}_{\tilde l,n,\tilde m}$ as~\eqref{eq:2D_ss}.
\STATE Perform Algorithm 1 to obtain ${\hat{  \boldsymbol{\kappa}}}_i^0=\left[{\hat \kappa _{\theta _i^0}} ,{\hat \kappa _{v_i^0}}\right]^T$ for $i = 1,2, \ldots,U$.
    \STATE Perform Algorithm 2 to obtain $({\hat \theta _i},{\hat v_i})$'s.
        \STATE Replace $(\theta,v)$ with $(\theta,r)$ and $(r,v)$ and repeat steps 1-3 in parallel manner to obtain $({\hat \theta _j},{\hat v_j})$ and $({\hat \theta _k},{\hat v_k})$ for $j,k = 1,2, \ldots,U$.
         \STATE Re-match and obtain the 3D-parameter estimations $({\hat \theta ^\star _i},{\hat r ^\star_i},{\hat v ^\star_i})$ for $i=1,2,...,U$ as \eqref{eq:pair} .
		\ENSURE optimal parameter pairing $({\hat \theta  ^\star_i},{\hat r ^\star_i},{\hat v ^\star_i})$'s.  
    \end{algorithmic}
\end{algorithm}
\vspace{-0.5cm}
\subsection{Analyze for Computational Complexity}\label{subsec: complexity}
Let $\tilde Z_1>\tilde Z_2>\tilde Z_3$ be a decreasing sort of $\tilde L$, $\tilde N$ and $\tilde M$. The spatially-smoothed sub-signal observation sizes $\tilde Z_1$, $\tilde Z_2$ and $\tilde Z_3$ are corresponding to the original signal observation sizes of $Z_1$, $Z_2$ and $Z_3$, respectively. For the proposed PI-2DMUSIC algorithm, its complexity comes from the covariance-matrix calculation, EVD and LM-based spatial spectrum optimization, whose complexities are ${\rm O}\left( {\tilde Z_1^3\tilde Z_2^3} \right)$, ${\rm O}\left( {\tilde Z_1^2\tilde Z_2^2} S_3 \right)$ and
${\rm O}\left( {{{\tilde Z_1^2\tilde Z_2^2}}U\log \log \left( {{\varepsilon ^{ - 1}}} \right)} \right)$, respectively,
with the error tolerance $\varepsilon$ and the notation $S_3=Z_3-\tilde Z_3+1$. Since it typically holds that $U\log \log \left( {{\varepsilon ^{ - 1}}} \right)\ll \min \left\{{\tilde Z_1\tilde Z_2},S_3\right\}$, the main complexity of the proposed algorithm comes from only the covariance-matrix calculation and EVD. Thus, the computational complexity of the proposed PI-2DMUSIC algorithm and existing super-resolution JARVE algorithms are summarized in the following TABLE I.
\begin{table}[htbp]
\caption{Algorithm Complexity Comparison}
\begin{center}
\begin{tabular}{ | c| c|} 
\hline\xrowht{10pt}
{\bf Algorithm} &{\bf Complexity} \\ 
\hline
PI-2DMUSIC&${\rm O}\left( {\tilde Z_1^3\tilde Z_2^3}  \right)+{\rm O}\left( {\tilde Z_1^2\tilde Z_2^2} S_3 \right)$\\
\hline
3D-MUSIC~\cite{3D-MUSIC} &${\rm O}\left( {{{\tilde L^2 \tilde N^2 \tilde M^2}}}{G_\theta }{G_r}{G_v} \right)$\\
\hline	
3D-ESPRIT~\cite{3D-ESPRIT-hu} &${\rm O}\left( {{{ {\tilde L^3\tilde N^3 \tilde M^3}}}} \right)+{\rm O}\left( {{{ {\tilde L^2 \tilde N^2 \tilde M^2}}S}} \right)$\\
\hline
\end{tabular}
\end{center}
\label{table1}
\end{table}

In Table I, the notations ${G_\theta },{G_r},{G_v}$ denote the number of 3D-MUSIC search grids of azimuth, range and velocity, respectively, which is usually extremely large for high-precision estimation; the number of spatially-smoothed snapshots $S=(L-\tilde L+1)(N-\tilde N+1)(M-\tilde M+1)$. Since it typically holds that ${\tilde Z_1\tilde Z_2} \ll {\tilde L\tilde N\tilde M} \ll {G_\theta }{G_r}{G_v}$, and $S_3 \ll S \ll {G_\theta }{G_r}{G_v}$, the complexity of proposed PI-2DMUSIC algorithm is much lower than the 3D-MUSIC algorithm~\cite{3D-MUSIC} and 3D-ESPRIT algorithm~\cite{3D-ESPRIT-hu}.

\vspace{-0.1cm}
\section{Numerical Results}\label{sec:simulation}
In this section, we numerically evaluate the RMSE performance and computational complexity of the proposed PI-2DMUSIC algorithm and existing JARVE algorithms. To make the evaluation more practical, we take the 5G NR reference signals as sensing signals, and use the system parameters in the following TABLE II.
\vspace{-0.1cm}									
\begin{table}[h!]
\caption{System Parameters}
\begin{center}
\begin{tabular}{ | c| c|} 
\hline\xrowht{10pt}
{\bf Parameter} &{\bf Value} \\ 
\hline
Carrier frequency ${f_{\rm{c}}}$&25G Hz\\
\hline	
Subcarrier spacing $\Delta f$&120 kHz\\
\hline
OFDM Data duration $T$&8.33 $\mu$s \\
\hline
OFDM CP duration $T_{{\rm{cp}}}$&0.59 $\mu$s\\
\hline	
Number of receive antennas $L$& [8:2:36]\\
\hline
Number of subcarriers $N$& [88:10:138]\\
\hline	
Number of OFDM symbols $M$& [48:8:88]\\
\hline	
Number of experiments $W$&200\\
\hline
\end{tabular}
\end{center}
\label{table2}
\end{table}


\vspace{-0.15cm}	
The RMSE for each parameter ${\boldsymbol{\mu }} = {\boldsymbol{\theta }},{\bf{r}}$ and ${\bf{v}}$, respectively, are defined as 
\begin{align}
RMSE({\boldsymbol{\mu }}){\rm{ }} \buildrel \Delta \over = \sqrt {\frac{1}{W}\sum\limits_{l = 1}^W {\frac{{\parallel {\boldsymbol{\mu }} - {\boldsymbol{\hat \mu }}\parallel _2^2}}{U}} }.
\end{align}

\vspace{-0.3cm}	
\subsection{Estimation Performance Comparison}\label{RMSE}
In this subsection, the estimation performance of both the proposed PI-2DMUSIC algorithm and existing algorithms are evaluated in terms of RMSE. The details of the existing 3D-DFT, 3D-MUSIC, and 3D-ESPRIT algorithms are referred to~\cite{3D-DFT,3D-MUSIC} and~\cite{3D-ESPRIT-hu}, respectively, which are omitted herein for brevity. The number of sensing targets is set as $U = 3$, and the targets' azimuth, range and velocity parameters are set as ($20^\circ$, 39.73m, -10m/s), ($-23.16^\circ$, 60.5m, 29.61m/s) and ($-10.6^\circ$, 80.21m, 10.11m/s), respectively. For fair comparison, the sizes of the spatially-smoothed sub-signal observations for all algorithms are fixed as $\tilde N = 40,\tilde M =25,\tilde L = 6$. 


\subsubsection{RMSE versus SNR}\label{subsubsec: RMSE_SNR}
\begin{figure}[H]
\centering
\subfloat[Azimuth-estimation RMSE vs SNR]{
		\includegraphics[width=.99\columnwidth,height=.74\columnwidth]{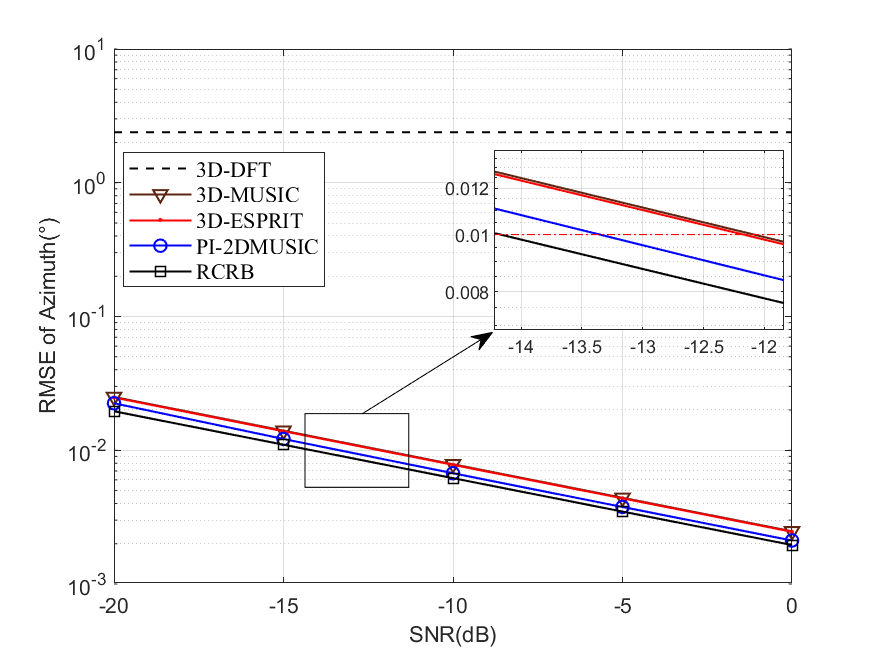}}\\
\subfloat[Range-estimation RMSE vs SNR]{
		\includegraphics[width=.99\columnwidth,height=.73\columnwidth]{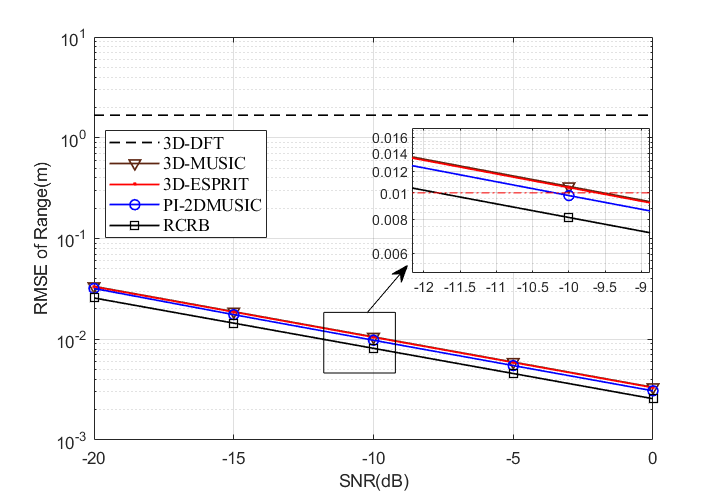}}\\
\subfloat[Velocity-estimation RMSE vs SNR]{
		\includegraphics[width=.99\columnwidth,height=.73\columnwidth] {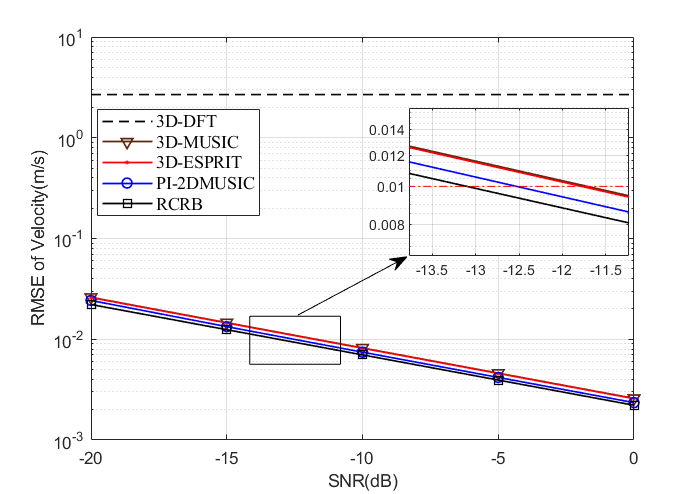}}\\
\caption{Comparison of JARVE RMSE versus SNR.}
\label{Fig4}
\end{figure}

Fix $L=16$, $N=128$ and $M=80$. Fig.~\ref{Fig4} (a) plots the RMSE for azimuth estimation versus the SNR. For each super-resolution algorithm including the proposed PI-2DMUSIC and the existing 3D-MUSIC together with 3D-ESPRIT algorithms, the RMSE decreases monotonically as the SNR increases; while for the traditional 3D-DFT algorithm, the RMSE almost keeps constant around 2.4 degree in the SNR interval between -20dB and 10dB, due to the limited antenna array aperture. The proposed PI-2DMUSIC algorithm improves the RMSE performance compared to the 3D-MUSIC and 3D-ESPRIT algorithms, and suffers from very slight RMSE performance degradation compared to the root of CRB, which is abbreviated as RCRB hereby. Specifically, at an RMSE level of $10^{-2}$ degree, the proposed algorithm achieves 1.25dB SNR improvement compared to the existing algorithms, and suffers from only 0.65dB SNR degradation compared to the RCRB.

Fig.~\ref{Fig4} (b) plots the range-estimation RMSE versus the SNR. For each super-resolution algorithm, the RMSE decreases as the SNR increases; while for the traditional 3D-DFT algorithm, the RMSE almost keeps constant around 1.7 meters, due to the limited range estimation Rayleigh resolution~\cite{3D-DFT} calculated as $\Delta r=c/(2B)\approx9.77$m. The proposed PI-2DMUSIC algorithm improves the RMSE performance compared to the benchmark 3D-MUSIC and 3D-ESPRIT algorithms, and suffers from very slight RMSE performance degradation compared to the root of CRB. Specifically, at an RMSE level of $10^{-2}$ meters, the proposed algorithm achieves 0.7dB SNR improvement compared to the existing algorithms, and suffers from only 1.5dB SNR degradation compared to the RCRB.

Fig.~\ref{Fig4} (c) illustrates the velocity-estimation RMSE versus the SNR. The RMSE for each super-resolution algorithm decreases as the SNR increases, but the RMSE for the traditional 3D-DFT algorithm almost keeps constant around 2.7 m/s, due to the limited velocity estimation Rayleigh resolution~\cite{3D-DFT} calculated as $\Delta v=c/(2f_{\rm{c}}M\bar T)\approx 8.4$m/s. The proposed PI-2DMUSIC algorithm improves the RMSE performance compared to the benchmark 3D-MUSIC and 3D-ESPRIT algorithms, and suffers from very slight RMSE performance degradation compared to the root of CRB. Specifically, at an RMSE level of $10^{-2}$ m/s, the proposed algorithm achieves 0.75dB SNR improvement compared to the existing algorithms, and suffers from only 0.5dB SNR degradation compared to the RCRB. 

From Fig.~\ref{Fig4}, it can be observed that the 3D-ESPRIT algorithm achieves almost the same RMSE performance as the 3D-MUSIC algorithm. Since the 3D-MUSIC algorithm has extremely high even unaffordable computational complexity, we take the 3D-ESPRIT algorithm as the existing benchmark algorithm in the latter simulations. Also, the 3D-DFT algorithm will not be compared in the latter simulations, due to its worse RMSE performance illustrated in Fig.~\ref{Fig4}.  

\subsubsection{RMSE versus Number of Antennas, Subcarriers and Symbols}\label{subsubsec: RMSE_afv}
\begin{figure}[H]
\centering
\subfloat[Azimuth-estimation RMSE vs Number of Antennas $L$]{
		\includegraphics[width=.99\columnwidth,height=.74\columnwidth]{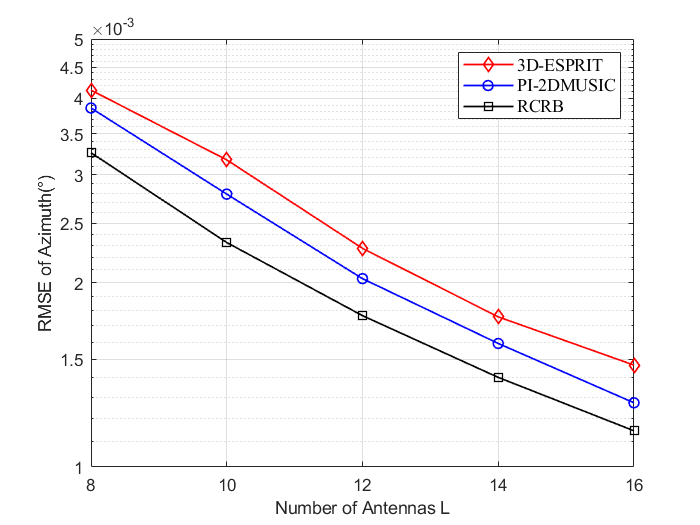}}\\
\subfloat[Range-estimation RMSE vs Number of Subcarriers $N$]{
		\includegraphics[width=.99\columnwidth,height=.73\columnwidth] {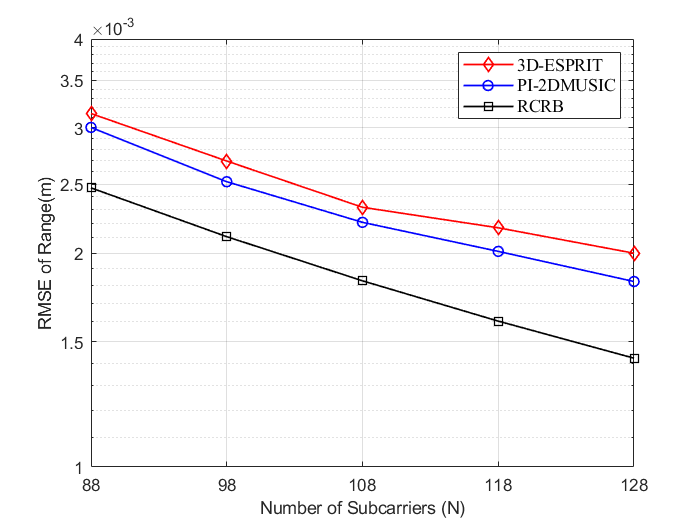}}\\
\subfloat[Velocity-estimation RMSE vs Number of Symbols $M$]{
		\includegraphics[width=.99\columnwidth,height=.73\columnwidth]{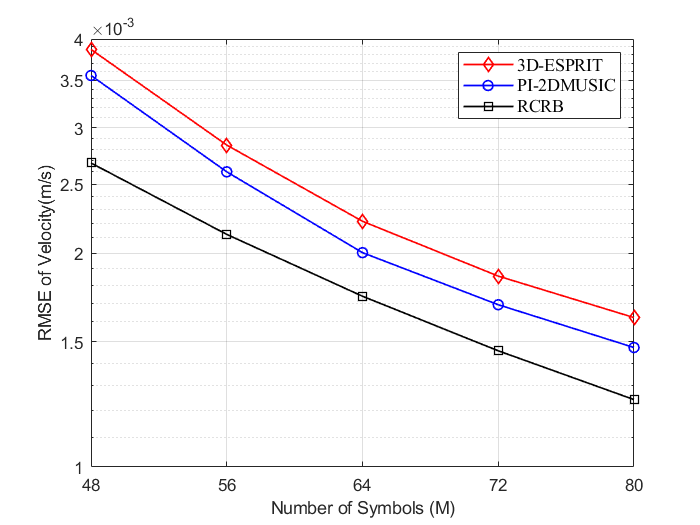}}\\
\caption{Comparison of JARVE RMSE versus Number of Antennas, Subcarriers and Symbols for SNR=5dB.}
\label{Figafv}
\end{figure}

Fig.~\ref{Figafv} (a) plots the effect of the number of receive antennas $L$ on the RMSE for azimuth estimation
by the fixed SNR=5dB, $N$=128 and $M$=80. In general, the 
RMSE descends as $L$ increases. The proposed PI-2DMUSIC algorithm achieves more significant RMSE performance improvement for larger $L$, compared to the benchmark 3D-ESPRIT algorithm. Specifically, for $L=16$, the proposed algorithm can decrease the azimuth-estimation RMSE by $15.5\%$.

Fig.~\ref{Figafv} (b) demonstrates the effect of the number of receive subcarriers $N$ on the RMSE of range estimation for fixed SNR=5dB, $L=16$ and $M=80$. Briefly, the RMSE descends as $N$ increases. The proposed PI-2DMUSIC algorithm achieves more significant RMSE performance improvement for larger $N$, compared to the 3D-ESPRIT algorithm. Specifically, for $N=128$, the proposed algorithm can decrease the range-estimation RMSE by $9.6\%$.

Fig.~\ref{Figafv} (c) illustrates the RMSE for velocity estimation versus the number of symbols $M$ by fixed SNR=5dB, $L=16$ and $N=128$.  Generally, the RMSE descends as $M$ increases. Compared to the benchmark 3D-ESPRIT algorithm, the proposed PI-2DMUSIC algorithm achieves more significant RMSE performance improvement for larger $M$. Specifically, for fixed $M=80$, the proposed algorithm can decrease the velocity-estimation RMSE by and $7.7\%$.

\subsection{Computational Complexity Comparison}
In this subsection, the computational complexity is compared in terms of the consumed average time of executing each algorithm once. The spatially-smoothed signal observation sizes are fixed as $\tilde L=3L/4, \tilde N=N/10$ and $\tilde M=3M/8$. In order to avoid the extremely-high complexity of 3D-MUSIC algorithm without affecting the estimation accuracy, the search grid of 3D-MUSIC algorithm is taken as a quarter of the RMSE of 3D-ESPRIT algorithm. All results are performed on the MATLAB R2022b software in the computer platform with 64-bit Windows 10 operating system, 32 GB random-access memory and a Core-i9 central processing unit at 3 GHz. 

\begin{figure}[H]
\centering
\subfloat[Computational Time vs Number of Antennas $L$]{
		\includegraphics[width=.99\columnwidth] {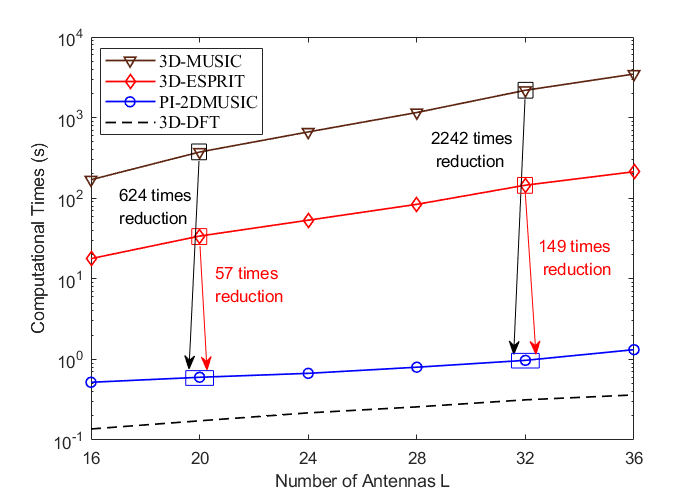}}\\

\subfloat[Computational Time vs Number of Subcarriers $N$]{
		\includegraphics[width=.99\columnwidth] {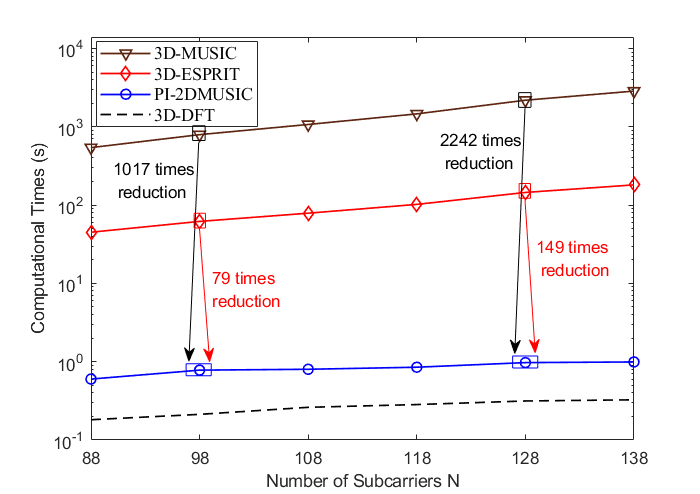}}\\


\subfloat[Computational Time vs Number of Symbols $M$]{
		\includegraphics[width=.99\columnwidth] {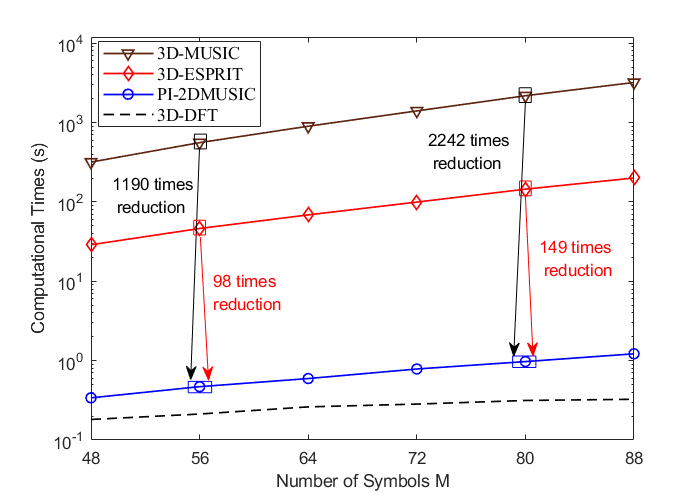}}\\
\caption{Comparison of Computational Time versus Number of Antennas, Subcarriers and Symbols.}
\label{Figtime}
\end{figure}

Fig.~\ref{Figtime} (a) demonstrates the computational time versus the number of antennas $L$ for fixed $N=128$ and $M=80$. The computational time of the proposed PI-2DMUSIC algorithm is much shorter than the 3D-ESPRIT and the 3D-MUSIC super-resolution algorithms. Furthermore, the complexity reduction of the proposed algorithm is more significant for larger $L$. Specifically, for $L=20$, the executing time of the proposed algorithm is 0.6s, which is approximately 57 times shorter than the 33.9s consumed by the 3D-ESPRIT algorithm, and 624 times shorter than the 374.3s consumed by the 3D-MUSIC algorithm. For a larger $L=32$, the computational time for the proposed PI-2DMUSIC algorithm is approximately 149 times shorter than the 3D-ESPRIT algorithm and 2242 times shorter than the 3D-MUSIC algorithm. 

Fig.~\ref{Figtime} (b) plots the computational time versus the number of subcarriers $N$ for fixed $L=32$ and $M=80$. Briefly, the computational time of the proposed PI-2DMUSIC
algorithm is much shorter than the 3D-ESPRIT and the 3D-MUSIC super-resolution algorithms. Furthermore, the complexity reduction of the proposed algorithm is more significant for larger $N$. Specifically, as $N$ increases from 98 to 128, the executing time reduction of the proposed algorithm is from 79 times to 149 times compared with the 3D-ESPRIT algorithm, and is from 1017 times to 2242 times compared with the 3D-MUSIC algorithm.


Fig.~\ref{Figtime} (c) illustrates the computational time versus the number of symbols $M$ for fixed $L=32$ and $N=128$. In
general, the proposed PI-2DMUSIC algorithm can achieve computational time reduction by two orders of magnitude and three orders of magnitude compared to the 3D-ESPRIT and the 3D-MUSIC algorithms, respectively. Furthermore, the complexity reduction of the proposed algorithm is more significant for larger $M$. Specifically, as $M$ increases from 56 to 88, the executing time reduction of the proposed algorithm is from 98 times to 149 times compared with the 3D-ESPRIT algorithm, and is from 1190 times to 2242 times compared with the 3D-MUSIC algorithm.

From Fig.~\ref{Figtime}, although the computational times of the 3D-DFT algorithm is least, its estimation accuracy is poor as numerically shown in previous Fig.~\ref{Fig4}.



\section{Conclusions}\label{conslusion}
This paper has investigated the joint azimuth-range-velocity estimation for an OFDM-based ISAC systems. The CRBs for the joint 3D-parameter estimation are derived to provide the performance bound. A PI-2DMUSIC estimation algorithm is proposed to achieve both extremely low complexity and super resolution. With the practical parameters of 5G New Radio, simulation results verify that the proposed algorithm can achieve significant complexity reduction and obvious estimation performance improvement. For instance, for the case of 32 receive antennas, 128 subcarriers and 80 symbols, the proposed algorithm can reduce the computational complexity by 2242 times compared to the existing 3D-MUSIC super-resolution algorithm, and by 149 times compared to the existing 3D-ESPRIT algorithm. More importantly, the complexity reduction and estimation performance improvement of proposed algorithm becomes more obvious for larger SNR and more sensing resources in the space-frequency-time domains (i.e, number of antennas, subcarriers and symbols). As for future work, the multi-parameter joint estimation in complex multi-path channels can be further studied, and the multiple-BS cooperative sensing is also promising to improve the sensing performance especially in the cell-edge areas.


\begin{appendix}



\subsection{Derivation of Fisher Information Matrix}\label{app:F}
Before deriving the Fisher information matrix, we give the following equation result for any matrix ${\bf{U}}$ and ${\bf{V}}$
\begin{align}\label{eq:result1}
{\mathop{\rm Re}\nolimits} \left\{ {\bf{U}} \right\}{\left( {{\mathop{\rm Re}\nolimits} \left\{ {\bf{V}} \right\}} \right)^T} = \frac{1}{2}\left( {{\mathop{\rm Re}\nolimits} \left\{ {{\bf{U}}{{\bf{V}}^H}} \right\} + {\mathop{\rm Re}\nolimits} \left\{ {{\bf{U}}{{\bf{V}}^T}} \right\}} \right),\nonumber\\
{\mathop{\rm Im}\nolimits} \left\{ {\bf{U}} \right\}{\left( {{\mathop{\rm Im}\nolimits} \left\{ {\bf{V}} \right\}} \right)^T} = \frac{1}{2}\left( {{\mathop{\rm Re}\nolimits} \left\{ {{\bf{U}}{{\bf{V}}^H}} \right\} - {\mathop{\rm Re}\nolimits} \left\{ {{\bf{U}}{{\bf{V}}^T}} \right\}} \right),\nonumber\\
{\mathop{\rm Re}\nolimits} \left\{ {\bf{U}} \right\}{\left( {{\mathop{\rm Im}\nolimits} \left\{ {\bf{V}} \right\}} \right)^T} = \frac{1}{2}\left( {{\mathop{\rm Im}\nolimits} \left\{ {{\bf{U}}{{\bf{V}}^T}} \right\} - {\mathop{\rm Im}\nolimits} \left\{ {{\bf{U}}{{\bf{V}}^H}} \right\}} \right).
\end{align}


From~\eqref{eq:result1}, the block sub-matrices of Fisher information matrix in~\eqref{eq:Fisher matrix} can be expressed as
\allowdisplaybreaks[3]
\begin{align}\label{eq:APPFij11}
{{\bf{F}}_{11}} &= {\mathbb E}\left\{ {\frac{{\partial \ell ({\bf{z}}|{\bf{\Gamma}})}}{{\partial {{\bf{\Gamma }}_1}}}{{\left( {\frac{{\partial \ell ({\bf{z}}|{\bf{\Gamma}})}}{{\partial {{\bf{\Gamma }}_1}}}} \right)}^T}} \right\}\nonumber\\
 &= \frac{2}{{{\sigma ^4}}}{\mathop{\rm Re}\nolimits} \left\{ {{{\bf{\Sigma }}^H}{{\bf{\dot A}}^H} {\mathbb E}\left\{ {{\boldsymbol{\omega}}{{\boldsymbol{\omega}}^H}} \right\}{\bf{\dot A\Sigma }}} \right\}\nonumber\\
 &\quad + \frac{2}{{{\sigma ^4}}}{\mathop{\rm Re}\nolimits} \left\{ {{{\bf{\Sigma }}^H}{{\bf{\dot A}}^H}{\mathbb E}\left\{ {{\boldsymbol{\omega}}{{\boldsymbol{\omega}}^T}} \right\}{{\bf{\dot A}}^*}{{\bf{\Sigma }}^*}} \right\}\nonumber\\
 &\buildrel (a) \over = \frac{2}{{{\sigma ^2}}}{\mathop{\rm Re}\nolimits} {\{ {{\bf{\Sigma }}^H}{{\bf{\dot A}}^H}{\bf{\dot A\Sigma }}\} },
\end{align}
where the equation (a) comes from the facts that ${\mathbb E}\left\{ {{\boldsymbol{\omega}}{{\boldsymbol{\omega}}^H}} \right\}=\sigma^2 \bf I$ and ${\mathbb E}\left\{ {{\boldsymbol{\omega}}{{\boldsymbol{\omega}}^T}} \right\}=0$.

Similarly, we have the following block sub-matrices of Fisher information matrix
\begin{align}\label{eq:APPFij120}
{{\bf{F}}_{1{{\bf{\alpha }}^{\rm r}}}} &= {\mathbb E}\left\{ {\frac{{\partial \ell ({\bf{z}}|{\bf{\Gamma}})}}{{\partial {{\bf{\Gamma }}_1}}}{{\left( {\frac{{\partial \ell ({\bf{z}}|{\bf{\Gamma}})}}{{\partial {{\boldsymbol{\alpha }}^{\rm r}}}}} \right)}^T}} \right\}\nonumber\\
 &= \frac{2}{{{\sigma ^2}}}{\mathop{\rm Re}\nolimits} \{ {{\bf{\Sigma }}^H}{{\bf{\dot A}}^H}{\bf{A}}\},\nonumber\\
{{\bf{F}}_{1{{\bf{\alpha }}^{\rm i}}}} &= {\mathbb E}\left\{ {\frac{{\partial \ell ({\bf{z}}|{\bf{\Gamma}})}}{{\partial {{\bf{\Gamma }}_1}}}{{\left( {\frac{{\partial \ell ({\bf{z}}|{\bf{\Gamma}})}}{{\partial {{\boldsymbol{\alpha }}^{\rm i}}}}} \right)}^T}} \right\}\nonumber\\
 &=  - \frac{2}{{{\sigma ^2}}}{\mathop{\rm Im}\nolimits} \{ {{\bf{\Sigma }}^H}{{\bf{\dot A}}^H}{\bf{A}}\},
\end{align}
which yields the following Fisher information matrix as in~\eqref{eq:F12} 
 \begin{align}\label{eq:APPFij12}
{{\bf{F}}_{12}} &= \left[ {{{\bf{F}}_{1{{\bf{\alpha }}^{\rm r}}}},{{\bf{F}}_{1{{\bf{\alpha }}^{\rm i}}}}} \right]\nonumber\\
 &= \frac{2}{{{\sigma ^2}}}\left[ {{\mathop{\rm Re}\nolimits} \{ {{\bf{\Sigma }}^H}{{\bf{\dot A}}^H}{\bf{A}}\} , - {\mathop{\rm Im}\nolimits} \{ {{\bf{\Sigma }}^H}{{\bf{\dot A}}^H}{\bf{A}}\} } \right].
\end{align}
\allowdisplaybreaks[1]

Also, the following block sub-matrices of Fisher information matrix still hold
\begin{align}\label{eq:APPFij220}
{{\bf{F}}_{{{\bf{\alpha }}^{\rm r}}{{\bf{\alpha }}^{\rm r}}}} &= {\mathbb E}\left\{ {\frac{{\partial \ell ({\bf{z}}|{\bf{\Gamma}})}}{{\partial {{\bf{\alpha }}^{\rm r}}}}{{\left( {\frac{{\partial \ell ({\bf{z}}|{\bf{\Gamma}})}}{{\partial {{\bf{\alpha }}^{\rm r}}}}} \right)}^T}} \right\}\nonumber\\
 &= \frac{2}{{{\sigma ^2}}}{\mathop{\rm Re}\nolimits} \left\{ {{{\bf{A}}^H}{\bf{A}}} \right\},\nonumber\\
{{\bf{F}}_{{{\bf{\alpha }}^{\rm r}}{{\bf{\alpha }}^{\rm i}}}} &= {\mathbb E}\left\{ {\frac{{\partial \ell ({\bf{z}}|{\bf{\Gamma}})}}{{\partial {{\bf{\alpha }}^{\rm r}}}}{{\left( {\frac{{\partial \ell ({\bf{z}}|{\bf{\Gamma}})}}{{\partial {{\bf{\alpha }}^{\rm i}}}}} \right)}^T}} \right\}\nonumber\\
 &=- \frac{2}{{{\sigma ^2}}}{\mathop{\rm Im}\nolimits} \left\{ {{{\bf{A}}^H}{\bf{A}}} \right\},\nonumber\\
{{\bf{F}}_{{{\bf{\alpha }}^{\rm i}}{{\bf{\alpha }}^{\rm r}}}} &= {\bf{F}}_{{{\bf{\alpha }}^{\rm r}}{{\bf{\alpha }}^{\rm i}}}^T \buildrel (a) \over= \frac{2}{{{\sigma ^2}}}{\mathop{\rm Im}\nolimits} \left\{ {{{\bf{A}}^H}{\bf{A}}} \right\},
\nonumber\\
{{\bf{F}}_{{{\bf{\alpha }}^{\rm i}}{{\bf{\alpha }}^{\rm i}}}} &= {\mathbb E}\left\{ {\frac{{\partial \ell ({\bf{z}}|{\bf{\Gamma}})}}{{\partial {{\bf{\alpha }}^{\rm r}}}}{{\left( {\frac{{\partial \ell ({\bf{z}}|{\bf{\Gamma}})}}{{\partial {{\bf{\alpha }}^{\rm i}}}}} \right)}^T}} \right\}\nonumber\\
 &= \frac{2}{{{\sigma ^2}}}{\mathop{\rm Re}\nolimits} \left\{ {{{\bf{A}}^H}{\bf{A}}} \right\},
 \end{align}
where (a) comes from the fact that the imaginary part of the Hermitian matrix $  {\bf A}^H {\bf A}$ is skew-symmetric. From \eqref{eq:APPFij220}, the following Fisher information matrix as in~\eqref{eq:F22} is obtained  
 \begin{align}\label{eq:APPFij22}
{{\bf{F}}_{22}} &= {\mathbb E}\left\{ {\frac{{\partial \ell ({\bf{z}}|{\bf{\Gamma}})}}{{\partial {{\bf{\Gamma }}_2}}}{{\left( {\frac{{\partial \ell ({\bf{z}}|{\bf{\Gamma}})}}{{\partial {{\bf{\Gamma }}_2}}}} \right)}^T}} \right\}\nonumber\\
 &= \left[ {\begin{array}{*{20}{c}}
{{{\bf{F}}_{{{\bf{\alpha }}^{\rm r}}{{\bf{\alpha }}^{\rm r}}}}}&{{{\bf{F}}_{{{\bf{\alpha }}^{\rm r}}{{\bf{\alpha }}^{\rm i}}}}}\nonumber\\
{{{\bf{F}}_{{{\bf{\alpha }}^{\rm i}}{{\bf{\alpha }}^{\rm r}}}}}&{{{\bf{F}}_{{{\bf{\alpha }}^{\rm i}}{{\bf{\alpha }}^{\rm i}}}}}
\end{array}} \right]\nonumber\\
 &= \frac{2}{{{\sigma ^2}}}\left[ {\begin{array}{*{20}{c}}
{{\mathop{\rm Re}\nolimits} \left\{ {{{\bf{A}}^H}{\bf{A}}} \right\}}&{ - {\mathop{\rm Im}\nolimits} \left\{ {{{\bf{A}}^H}{\bf{A}}} \right\}}\\
{{\mathop{\rm Im}\nolimits} \left\{ {{{\bf{A}}^H}{\bf{A}}} \right\}}&{{\mathop{\rm Re}\nolimits} \left\{ {{{\bf{A}}^H}{\bf{A}}} \right\}}
\end{array}} \right].
\end{align}

 
\subsection{CRB Derivation for Single-Target JARVE}\label{app:CRBs}
For the case of single target, the manifold matrix in~\eqref{eq:AA123} is given by ${\bf{A}} ={{\bf{a}}}\left( {{\theta },{r},{v}} \right)= {{{\bf{a}}}_\theta }\otimes{{{\bf{a}}}_r }\otimes{{{\bf{a}}}_v }$.
Similarly, the partial derivative of manifold matrix in~\eqref{eq:diff A} is given by ${\bf{\dot A}} = \left[ {{{{\bf{\dot a}}}_\theta },{{{\bf{\dot a}}}_r},{{{\bf{\dot a}}}_v}} \right]\in {{\mathbb C}^{LMN \times 3}}$. And the relationship between the partial derivative of rotating vector and rotating vector can be written as 
\begin{align}\label{eq:da2a} 
{{{\bf{\dot a}}}_\theta }\buildrel \Delta \over  = d{\bf{a}}/d\theta  = {{\bf{D}}_\theta }{\bf{a}},\nonumber\\
{{{\bf{\dot a}}}_r}\buildrel \Delta \over  = d{\bf{a}}/dr = {{\bf{D}}_r}{\bf{a}},\nonumber\\
{{{\bf{\dot a}}}_v} \buildrel \Delta \over = d{\bf{a}}/dr = {{\bf{D}}_v}{\bf{a}},
\end{align}
where the derivative matrices are given by
\begin{align}\label{eq:Da2a} 
{{\bf{D}}_\theta } &= {d_\theta }{\text {blkdiag}}\left\{ {{{\left[ {0,1,...,L - 1} \right]}^T}{{\bf{I}}_{NM}}} \right\},\nonumber\\
{{\bf{D}}_r} &= {d_r}{\text {blkdiag}}\left\{ {{{\bf{1}}_L} \otimes {{\left[ {0,1,...,N - 1} \right]}^T} \otimes {{\bf{1}}_M}} \right\},\nonumber\\
{{\bf{D}}_v} &= {d_v}{\text {blkdiag}}\left\{ {{{\bf{1}}_{LN}} \otimes {{\left[ {0,1,...,M - 1} \right]}^T}} \right\}.
\end{align}


Then, the CRBs of the interested parameters in \eqref{eq:CRBend} can be transformed into 
\begin{align}\label{eq:CRB_s1}
CRB({{\bf{\Gamma }}_1})&= \frac{{{\sigma ^2}}}{2}{\left[ { {\text{Re}}\{ ({\alpha ^H}{{{\bf{\dot A}}}^H}{\bf{P}}_{\bf{A}}^ \bot {\bf{\dot A}}\alpha )\} } \right]^{ - 1}},\nonumber\\
 &= \frac{{{{\left[ {{\text{Re}}\{  {\bf{J}}\} } \right]}^{ - 1}}}}{{2\gamma}},
 \end{align}
 where the orthogonal projection matrix ${\bf{P}}_{\bf{A}}^ \bot  = {\bf{I}} - {\bf{a}}{\left( {{{\bf{a}}^H}{\bf{a}}} \right)^{ - 1}}{{\bf{a}}^H} = {\bf{I}} - \frac{{\bf{a}}{{\bf{a}}^H}}{{LNM}}$, the SNR $\gamma=|\alpha|^2/{\sigma^2}$, and the matrix ${\bf{J}}$ is defined as follow
 \begin{align}
 {\bf{J}} &\buildrel \Delta \over= {{{\bf{\dot A}}}^H}{\bf{P}}_{\bf{A}}^ \bot {\bf{\dot A}}  \nonumber \\
 &= {{{\bf{\dot A}}}^H}{\bf{\dot A}} - \frac{1}{{LNM}}{{{\bf{\dot A}}}^H}{\bf{a}}{{\bf{a}}^H}{\bf{\dot A}} \label{eq:J0} \\
 &= \left[ {\begin{array}{*{20}{c}}
{{J_{11}}}&{{J_{12}}}&{{J_{13}}}\\
{{J_{21}}}&{{J_{22}}}&{{J_{23}}}\\
{{J_{31}}}&{{J_{32}}}&{{J_{33}}}
\end{array}} \right].\label{eq:J}
\end{align}

Substituting ${\bf{\dot A}} = \left[ {{{{\bf{\dot a}}}_\theta },{{{\bf{\dot a}}}_r},{{{\bf{\dot a}}}_v}} \right]\in {{\mathbb C}^{LMN \times 3}}$ into the two parts of~\eqref{eq:J0} yield, respectively, 
\begin{align}\label{eq:J12}
{{{\bf{\dot A}}}^H}{\bf{\dot A}} &= \left[ {\begin{array}{*{20}{c}}
{{\bf{\dot a}}_\theta ^H{{{\bf{\dot a}}}_\theta }}&{{\bf{\dot a}}_\theta ^H{{{\bf{\dot a}}}_r}}&{{\bf{\dot a}}_\theta ^H{{{\bf{\dot a}}}_v}}\\
{{\bf{\dot a}}_r^H{{{\bf{\dot a}}}_\theta }}&{{\bf{\dot a}}_r^H{{{\bf{\dot a}}}_r}}&{{\bf{\dot a}}_r^H{{{\bf{\dot a}}}_v}}\\
{{\bf{\dot a}}_v^H{{{\bf{\dot a}}}_\theta }}&{{\bf{\dot a}}_v^H{{{\bf{\dot a}}}_r}}&{{\bf{\dot a}}_v^H{{{\bf{\dot a}}}_v}}
\end{array}} \right],\nonumber\\
{{{\bf{\dot A}}}^H}{\bf{a}}{{\bf{a}}^H}{\bf{\dot A}} &= \left[ {\begin{array}{*{20}{c}}
{{\bf{\dot a}}_\theta ^H{\bf{a}}{{\bf{a}}^H}{{{\bf{\dot a}}}_\theta }}&{{\bf{\dot a}}_\theta ^H{\bf{a}}{{\bf{a}}^H}{{{\bf{\dot a}}}_r}}&{{\bf{\dot a}}_\theta ^H{\bf{a}}{{\bf{a}}^H}{{{\bf{\dot a}}}_v}}\\
{{\bf{\dot a}}_r^H{\bf{a}}{{\bf{a}}^H}{{{\bf{\dot a}}}_\theta }}&{{\bf{\dot a}}_r^H{\bf{a}}{{\bf{a}}^H}{{{\bf{\dot a}}}_r}}&{{\bf{\dot a}}_r^H{\bf{a}}{{\bf{a}}^H}{{{\bf{\dot a}}}_v}}\\
{{\bf{\dot a}}_v^H{\bf{a}}{{\bf{a}}^H}{{{\bf{\dot a}}}_\theta }}&{{\bf{\dot a}}_v^H{\bf{a}}{{\bf{a}}^H}{{{\bf{\dot a}}}_r}}&{{\bf{\dot a}}_v^H{\bf{a}}{{\bf{a}}^H}{{{\bf{\dot a}}}_v}}
\end{array}} \right].
\end{align}

By using the fact that ${{\bf{x}}^H}{\bf{y}} = {\text{Tr}}\left( {{\bf{y}}{{\bf{x}}^H}} \right)$ for any two column vectors $\bf{x}$ and $\bf{y}$, we can obtain that
\begin{align}\label{eq:j121}
{\bf{\dot a}}_\theta ^H{{{\bf{\dot a}}}_\theta } & = {\text{Tr}}\left( {{{\bf{D}}_\theta }{\bf{a}}{{\bf{a}}^H}{{\bf{D}}_\theta }} \right) \nonumber \\
&\buildrel (a) \over =  {\text{Tr}} \left( {{\bf{D}}_\theta ^2} \right)\nonumber\\
 &\buildrel (b) \over = d_\theta ^2LNM\frac{{(L - 1)(2L - 1)}}{6},
\end{align}
where (a) comes from the fact that all the diagonal elements of ${\bf{a}}{{\bf{a}}^H}$ are one, and (b) is from~\eqref{eq:Da2a}.

Similarly, we can obtain the following equations
 \begin{align}\label{eq:j122}
{\bf{\dot a}}_\theta ^H{\bf{a}}{{\bf{a}}^H}{{{\bf{\dot a}}}_\theta } 
 &= d_\theta ^2{\left( {LNM} \right)^2}\frac{{{{(L - 1)}^2}}}{4},\nonumber\\
{\bf{\dot a}}_\theta ^H{{{\bf{\dot a}}}_r} &= \frac{{\text{Tr}}\left( {{{\bf{D}}_\theta }} \right) {\text{Tr}} \left( {{{\bf{D}}_r}} \right)}{LNM},\nonumber\\
{\bf{\dot a}}_\theta ^H{\bf{a}}{{\bf{a}}^H}{{{\bf{\dot a}}}_r} 
 &= {\text{Tr}}\left( {{{\bf{D}}_\theta }} \right) {\text{Tr}}\left( {{{\bf{D}}_r}} \right).
 \end{align}

 Then, $J_{11}$ and $J_{12}$ in~\eqref{eq:J} can be obtained by using~\eqref{eq:J12},~\eqref{eq:j121} and~\eqref{eq:j122} as
 \begin{align}\label{eq:j11}
 {J_{11}} &= d_\theta ^2LNM({L^2} - 1)/12,\nonumber\\
 {J_{12}} &={\bf{\dot a}}_\theta ^H{{{\bf{\dot a}}}_r} - \frac{1}{{LNM}}{\bf{\dot a}}_\theta ^H{\bf{a}}{{\bf{a}}^H}{{{\bf{\dot a}}}_r} = 0.
 \end{align}
 
 Similarly, $J_{22}$, $J_{33}$ and ${J_{ij}},i,j = 1,2,3,i \ne j$, in~\eqref{eq:J} can be obtained as
\begin{align}\label{eq:j2233}
{J_{22}} &= d_r^2LNM({N^2} - 1)/12,\nonumber\\
{J_{33}} &= d_v^2LNM({M^2} - 1)/12,\nonumber\\
{J_{ij}} &= 0,i,j = 1,2,3,i \ne j.
\end{align}
By substituting~\eqref{eq:j11},~\eqref{eq:j2233} and~\eqref{eq:J} into~\eqref{eq:CRB_s1}, the single-target CRB of the interested parameters is given by
\begin{align}\label{eq:CRBs_end}
CRB({{\bf{\Gamma }}_1}) &= \frac{1}{{2\gamma}}{\left[ {\begin{array}{*{20}{c}}
{{J_{11}^{ - 1}}}&0&0\\
0&{{J_{22}^{ - 1}}}&0\\
0&0&{{J_{33}^{ - 1}}}
\end{array}} \right]}.
\end{align}

Finally, substituting~\eqref{eq:j11} and~\eqref{eq:j2233} into~\eqref{eq:CRBs_end} yields the equation~\eqref{eq:CRB123_s}, which completes the proof.
\end{appendix} 
\bibliography{IEEEabrv,my}
\bibliographystyle{IEEEtran}

\end{document}